\documentclass[report,12pt]{article}
\pdfoutput=1
\usepackage[utf8x]{inputenc}
\usepackage[T1]{fontenc}
\usepackage{multicol}
\usepackage{float}
\usepackage{vmargin}
\usepackage{amsmath}
\usepackage{amsfonts}
\usepackage{amssymb}
\usepackage{amsthm}
\usepackage{bbold}
\usepackage{latexsym}
\usepackage{feynmf}
\usepackage{color}
\usepackage{tensor}
\usepackage{graphics}
\usepackage{graphicx}
 \usepackage{pdfpages}
 \usepackage{tikz}
 \usetikzlibrary{positioning,calc}
 \usepackage{hyperref}
\usepackage{vmargin}
\setpapersize{A4}
  \setmargins{2.5cm}{1cm}    
	     {16cm}{25cm}    
	     {12pt}{25pt}    
	     {0pt}{30pt}     

\begin{document}
\thispagestyle{empty}
\begin{center}
{Reduction of couplings and its application in particle physics\\
	Finite theories\\
Higgs and top mass predictions}
\end{center}
\renewcommand{\thefootnote}{\alph{footnote}}
\begin{center}
Managing editor: Klaus Sibold\\
Authors:\\
Jisuke Kubo\footnote{Institute for Theoretical Physics, Kanazawa University, Kanazawa 902-1192, Japan\\
                     e-mail: jik at hep.s.kanazawa-u.ac.jp},
Sven Heinemeyer\footnote{Instituto de F\'isica de Cantabria (CSIC-UC), Santander, Spain\\
                     e-mail: Sven.Heinemeyer at cern.ch},
Myriam Mondrag\'on\footnote{Instituto de F\'isica, Universidad Nacional Aut\'onoma de M\'exico, M\'exico, M\'exico\\
                     e-mail: myriam at fisica.unam.mx},
Olivier Piguet\footnote{Dep. de F\'isica, Universidade Federal de Vi{\c c}osa, Brasil\\
                     e-mail: opiguet at yahoo.com}\\
Klaus Sibold\footnote{Institute for Theoretical Physics, Leipzig University, Leipzig, Germany\\
                     e-mail: sibold at physik.uni-leipzig.de},
Wolfhart Zimmermann\footnote{Max-Planck-Institut f\"ur Physik (Werner-Heisenberg-Institut),
		      D-80805 M\"unchen, Germany},
George Zoupanos\footnote{Physics Department, National Technical University, Athens, Greece\\
                     e-mail: George.Zoupanos at cern.ch}
\end{center}

\begin{abstract}
\noindent
In this report we tell the story of the notion {\sl reduction
of couplings} as we witnessed it in the course of time. Born
as an innocent child of renormalization theory it first
served the study of asymptotic behavior of several couplings
in a given model. Reduced couplings appeared as functions of a
primary one, compatible with the renormalization group equation
and thus solutions of a specific set of ordinary differential
equations. If these functions
have the form of power series the respective theories resemble
standard renormalizable ones and thus widen considerably the
area covered until then by symmetries as a tool for constraining
the number of couplings consistently. Still on the more abstract
level reducing couplings enabled one to construct theories with
$\beta$-functions vanishing to all orders of perturbation theory.
Reduction of couplings became physicswise truely
interesting and phenomenologically important when applied to the
standard model and its possible extensions. In particular in 
the context of supersymmetric theories it became the most
powerful tool known today once it was learned how to apply
it also to couplings having dimension of mass and to mass
parameters. Technically this all relies on the basic
property that reducing couplings is a renormalization scheme
independent procedure. Predictions of top and Higgs mass
prior to their experimental finding highlight the fundamental
physical significance of this notion. Twenty-two original
articles and one set of lectures are being commented, put
into historical perspective and interrelated with each other.\\
(Note: At
\href{http://pos.sissa.it/cgi-bin/reader/conf.cgi?confid=222}{Proceedings of Science, Sissa,
Trieste, Italy}  
the reader can find an electronic
version which includes the original articles.)
\end{abstract}


\pagebreak

\includepdf[pages=1,, scale=0.9,pagecommand={\thispagestyle{empty}},offset=20mm -20mm]{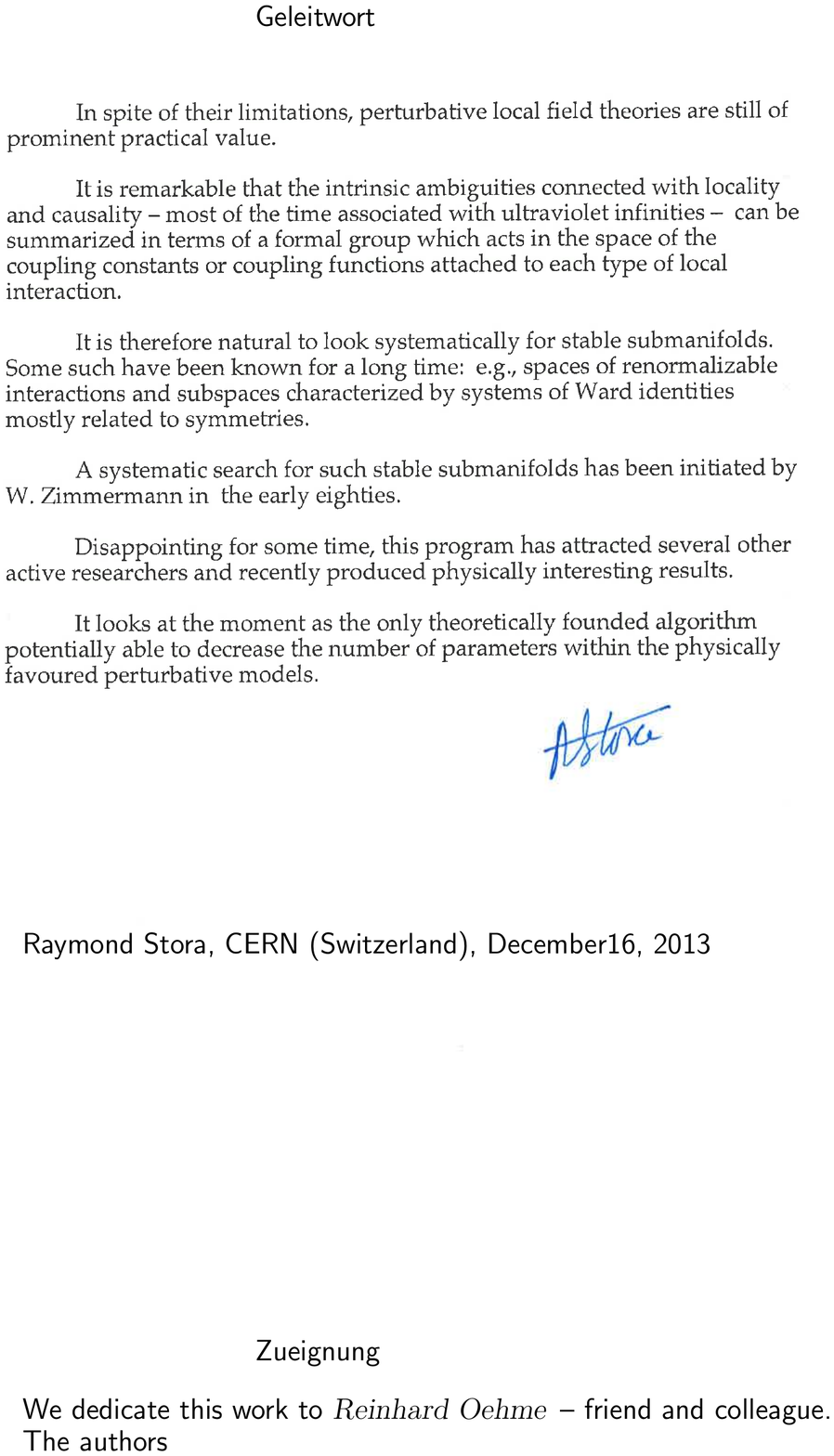}

\noindent
{\large List of papers treated in this work}\\

\noindent
W.\ Zimmermann\\
Reduction in the number of coupling parameters\\
Commun.\ Math.\ Phys.\ {\bf 97} (1985) 211-225\\

\noindent
R.\ Oehme, W.\ Zimmermann\\
Relation between effective couplings for asymptotically free models\\
Commun.\ Math.\ Phys.\ {\bf 97} (1985) 569-582\\

\noindent
R.\ Oehme, K.\ Sibold, W.\ Zimmermann\\
Renormalization group equations with vanishing lowest order of the primary $\beta$-function\\
Phys.\ Letts.\ {\bf B147} (1984)115-120\\

\noindent
R.\ Oehme, K.\ Sibold, W.\ Zimmermann\\
Construction of gauge theories with a single coupling parameter for Yang-Mills and
matter fields\\
Phys.\ Letts.\ {\bf B153} (1985)142-146\\

\noindent
J.\ Kubo, K.\ Sibold, W.\ Zimmermann\\
Higgs and top mass from reduction of couplings\\
Nucl.\ Phys.\ {\bf B259} (1985) 331-350\\

\noindent
K.\ Sibold, W.\ Zimmermann\\
Quark family mixing and reduction of couplings\\
Phys.\ Letts.\ {\bf B191} (1987) 427-430\\

\noindent
J.\ Kubo, K.\ Sibold, W.\ Zimmermann\\
New results in the  reduction of the standard model\\
Phys.\ Letts.\ {\bf B220} (1988) 185-191\\

\noindent
J.\ Kubo, K.\ Sibold, W.\ Zimmermann\\
Cancellation of divergencies and reduction of couplings
in the standard model\\
Phys.\ Letts.\ {\bf B220} (1989) 191-194\\

\noindent
J.\ Kubo\\
Precise determination of the top quark and Higgs masses
in the reduced standard theory for electroweak and strong interactions\\
Phys.\ Letts.\ {\bf B262} (1991) 472-476\\

\pagebreak
\noindent
Lucchesi, O. Piguet, K. Sibold\\
Vanishing $\beta-$functions in supersymmetric gauge theories\\
Helv.\ Phys.\ Acta {\bf 61} (1988) 321-344\\

\noindent
C.\ Lucchesi,\ O. Piguet, K.\ Sibold\\
Necessary and sufficient conditions for all order
vanishing $\beta-$functions in supersymmetric Yang-Mills
theories\\
Phys.\ Letts.\ {\bf B201} (1988) 241-244\\

O.\ Piguet, K.\ Sibold\\
Reduction of couplings in the presence of parameters\\
Phys.\ Letts.\ {\bf B 229} (1989) 83-89\\

\noindent
W.\ Zimmermann\\
Scheme independence of the reduction principle and asymptotic freedom in several
couplings \\
Commun.\ Math.\ Phys.\ {\bf 219} (2001) 221-245\\

\noindent
D.\ Kapetanakis, M.\ Mondrag\'on, G.\ Zoupanos\\
Finite unified models\\
Z.\ Phys.\ {\bf  C60} (1993) 181-186\\

\noindent
J.\ Kubo, M.\ Mondrag\'on, G.\ Zoupanos\\
Reduction of couplings and heavy top quark in the minimal SUSY GUT\\
Nucl.\ Phys.\ {\bf B424} (1994) 291-307\\

\noindent
J.\ Kubo, M.\ Mondrag\'on, G.\ Zoupanos\\
Perturbative unification of soft supersymmetry-breaking terms\\
Phys.\ Letts.\ {\bf B389} (1996) 523-532\\

\noindent
J.\ Kubo, M.\ Mondrag\'on, G.\ Zoupanos\\ 
Unification beyond GUTs: Gauge Yukawa unification\\
Acta\ Phys.\ Polon.\ {\bf B27} (1997) 3911-3944\\

\noindent
T.\ Kobayashi, J.\ Kubo, M.\ Mondrag\'on, G.\ Zoupanos\\
Constraints on finite soft supersymmetry-breaking terms\\
Nucl.\ Phys.\ {\bf  B511} (1998) 45-68\\ 

\noindent
T.\ Kobayashi, J.\ Kubo, G.\ Zoupanos\\
Further all loop results in softly broken supersymmetric gauge theories\\
Phys.\ Letts.\ {\bf B427} (1998) 291-299\\

\noindent
E.\ Ma, M.\ Mondrag\'on, G.\ Zoupanos\\
Finite $SU(N)^k$ unification\\
Journ.\ of High Energy Physics 0412 (2004) 026\\

\noindent
S.\ Heinemeyer, M.\ Mondrag\'on, G.\ Zoupanos\\
Confronting finite unified theories with low energy phenomenology\\
Journ.\ of High Energy Physics 0807 (2008) 135-164\\

\noindent
S.\ Heinemeyer, M.\ Mondrag\'on, G.\ Zoupanos\\
Finite theories after the discovery of a Higgs-like boson at the LHC\\
Phys.\ Letts.\ {\bf B718} (2013) 1430-1435\\

\noindent
M.\ Mondrag\'on, N.D.\ Tracas, G.\ Zoupanos\\
Reduction of Couplings in the MSSM\\ 
Phys.\ Letts.\ {\bf B728} (2014) 51-57\\

\pagebreak

\tableofcontents

\pagebreak

\section{Introduction} 
\noindent
Particle physics of today is well described by relativistic quantum
field theory (QFT) based on flat Minkowski spacetime. Comparison with
experiment works astonishingly well within the context of a gauge
theory based on the group $SU(3) \times SU(2) \times U(1)$, the
so called standard model (SM). Although quarks and gluons are
confined to form baryons and mesons a perturbative treatment of the
SM yields predictions which are in excellent agreement with
experiment and in practical terms one is able to separate quite well
the nonperturbative aspects from the perturbative ones. Similarly
gravitational effects do not yet seriously require to be considered
in particle physics although astrophysical results clearly point to 
the existence of dark matter and pose the ``missing mass'' problem
to which the SM does not give an answer. If one is interested
in the description of particle physics only, one may thus hand
over these fundamental problems to string theory,
quantum gravity or (non-commutative) extensions of spacetime and
study the SM and its extensions on flat spacetime in their own right.\\
It is precisely the outcome of such studies which we present
here in its historical context. The SM requires as input from experiment many
parameters: couplings, masses and mixing angles. Too many --
according to the taste of quite a few people -- to be considered
as being fundamental. Hence one calls for ideas to restrict the
number of parameters without spoiling the successes of the SM.
The two main principles which we invoke here are: reduction of
coupling parameters and finiteness. The first one relies on the
discovery that parameters which are a priori independent permit
ordering according to the degree with which they die out
when performing an asymptotic expansion for small coupling
in agreement with renomalization group equations. So, to be
specific, one may express ``secondary'' couplings as power series
in a ``primary'' one, hence the secondaries go together with
the primary to zero (obviously faster). The general solutions
of the renormalization group equations
for the secondaries are then found as deviations from their
power series in the primary coming with some arbitrary coefficient:
the integration constant which carries the information that
the secondary can also be an independent coupling in its own right.
But again: ``reduction'' works if this additional contribution
(which can depend on logarithms also) goes faster to zero than
the power series. The concept ``finiteness'' is most easily realized
in the context of supersymmetric theories and is in its truely
physical form understood as the vanishing of $\beta$-functions,
because those can be constructed as gauge parameter independent
quantities. Anomalous dimensions to the contrary are usually
gauge parameter dependent, hence only their gauge parameter
independent parts may be considered as physical and required to
vanish.\\
Although we stressed here the notion of reduction and finiteness
as convenient tools to search for a theoretically appealing and
experimentally satisfactory theory of particle physics it
is clear that they are interesting areas of research in their
own right.\\
The present report is not to be understood as a traditional
review paper, but rather as a guide to existing literature
in which these principles have been developped and fused to the
aim spelled out above: enriching the SM without loosing its
benefits. We therefore have first chosen those original papers
where the respective ideas have been worked out; then we put them
into a logical order (which is almost the same as time ordering)
and -- hopefully the most valuable contribution -- commented
them, in particular by relating them amongst each other.\\
The outline of the report is as follows. The papers of section 2 
introduce the notion of ``reduction of couplings''. In the examples
treated there it becomes in particular clear that a stability analysis of
(power series) solutions of the reduction equations is the appropriate tool
for embedding them in an enlightening neighbourhood. Many more
examples have been worked out, they can be found in reviews which
we quote. Section 3 is devoted to the application of the reduction
method to the SM. It turns out that a refined notion, called ``partial
reduction'', is needed in order to deal with the problem of
different asymptotic behavior (UV- versus IR-freedom) of the couplings.
It was possible to give either values or bounds to the Higgs
and top mass. In section 4 two topics are introduced: finiteness
in $N=1$ supersymmetric gauge theories and an extension of the
reduction method for including parameters carrying mass dimension
together with the proof that the reduction method is renormalization scheme
independent. Whereas the finiteness papers provide simple necessary
and sufficient criteria for vanishing $\beta$-functions operating
at one-loop order the other paper is crucial for correctly and
efficiently controlling all types of susy breaking needed later on.
Based on values of $\alpha_s$ etc.\ around 1990 reduction of couplings
in the SM eventually predicted for the Higgs mass roughly 65 GeV,
for the top mass roughly 100 GeV. Cancellation of quadratical
divergencies was already at the borderline of being compatible with
these numbers. Soon later precision experiments pointed towards
higher mass values. Trusting the reduction method, i.e.\ the
relevance of asymptotic expansions it was tempting to go one step
further and to ask for finiteness. Thus, section 5 has been devoted
to the development of this line of thought and some of its ramifications.
The key notion here became reduction of parameters carrying dimension.
It is based on the observation that also such parameters can give 
rise to closed renormalization orbits which can be found this way.\\
Still one remark for reading. Every section starts with an introduction
putting the subsections which consist of an original paper plus comment
into the respective context. Section 6 contains discussion and conclusions
for the whole set of papers.\\

\pagebreak

\section{Fundamentals: Asymptotic freedom, reduction of couplings}
{\sl Klaus Sibold}\\
In the context of $QCD$ an important property of the gauge coupling has 
been found: introducing an effective coupling which depends on the
characteristic energy scale of some process under consideration
it is seen that this coupling {\bf de}creases in strength when the energy
{\bf in}creases. So, for infinite energy the coupling vanishes and the
theory becomes free: this behaviour has been coined (UV-) asymptotic
freedom. This observation has first been made in the context of
perturbation theory but also non-perturbatively it played an
important role in the study of $QCD$.\\
It is then a natural question to ask in theories of more than one
coupling for a criterion that guarantees asymptotic freedom for all
couplings. This analysis has been performed by Zimmermann and Oehme
and lead Zimmermann by eliminating the running parameter in terms
of one -- the ``primary`` -- coupling to a set of ordinary differential
equations, the ``reduction equations``. Those are therefore to be
studied and solved. The special case of asymptotic freedom suggests
to demand that all couplings vanish together with the primary one
in the limit of weak coupling. One may hope
that the model being considered in perturbation theory has a non-perturbative
analogue to which it is a reasonable approximation.\\

\noindent
\subsection{Reduction in the number of coupling parameters}
Title: Reduction in the number of coupling parameters\\
Author: W.\ Zimmermann\\
Journal: \href{http://www.springerlink.com/index/QM27X33811641100.pdf}{Commun.\ Math.\ Phys.\ {\bf 97} (1985) 211-225}\\

Comment ({\sl Wolfhart Zimmermann}\,)  \\
The standard model of elementary particles involves a large number of 
parameters which are not constrained by any symmetry. Therefore, it is of 
considerable interest to find general concepts in quantum field theory which
can be used for reducing the number of independent parameters even in cases
where no suitable symmetry is available.\\
In the present work renormalizable models of quantum field theory are considered
which describe massless particles with an interaction given by several coupling
terms in the Lagrangian. A normalization mass is introduced for the purpose of
normalizing fields and defining finite coupling parameters. The renormalized Green's
functions of the model can be expanded as power series in the coupling parameters
at any given value of the normalization mass.\\
Field operators are normalized by their propagators at the normalization mass.
Coupling parameters are conveniently defined by specific values of appropriate
vertex functions at the normalization mass. The normalization mass is an auxiliary
parameter which may be chosen arbitrarily. A change of the normalization mass merely
implies a redefinition of fields and coupling parameters without affecting the model
as such. So the field operators are multiplied by positive factors. The coupling
parameters are modified by their defining vertex function at the new value of the
normalization mass. Thus an equivalent description of the model is obtained. These
equivalence transformations constitute the renormalization group under which the
system stays invariant.\\
The reduction principle proposed in this paper requires that all couplings can be
expressed as functions of one of them, the primary coupling, such that the resulting
system is again invariant under the renormalization group. Moreover, the following
requirements are imposed on the reduced couplings as functions of the primary one:\\

\begin{tabular}[t]{rl}
	(i)& The dependence should not involve the normalization mass,\\
       (ii)& in the weak coupling limit the reduced couplings should vanish\\
	   & together with the primary coupling,\\
      (iii)& the reduced couplings can be expanded with respect to powers\\
      	   &  of the primary coupling.
\end{tabular}      	     

\noindent
The first condition is obvious, since the normalization mass is only an auxiliary
parameter. Requirement (ii) also seems natural, but is already quite restrictive.
It cannot be imposed for many models. If the reduced model should resemble a
renormalizable theory, all couplings should have power series in the primary
coupling (requirement (iii)). Under this condition there is usually only a finite
number of solutions, if any.\\
Invariance under the the renormalization group leads to partial differential equations
for the Green's functions with respect to the couplings and the normalization mass.
Comparing these equations for the original and the reduced system one finds a set
of ordinary differential equations for the coupling parameters as functions of the
primary coupling. Its solutions should satisfy the requirements (i) -- (iii).
These are the reduction equations which form the basis for the studies in this 
work.\\
Any symmetry of a system by which all couplings can be expressed in terms of a single
one certainly leads to a solution of the reduction equations provided the symmetry
can be implemented in all orders of perturbation theory. In cases where a symmetry
cannot be established  in higher orders the reduction method may still lead to a
corresponding solution valid in all orders. But the main purpose of this work is to
provide the basis for finding reductions of a system which are not related to any 
symmetry.\\
An example is the Yukawa interaction of a spinor and a pseudoscalar field with a
quartic interaction of the pseudoscalar field in addition. Here the reduction
equation has a unique solution which expresses the coupling of the quartic interaction
as a function of the Yukawa coupling. No symmetry seems to be involved in this case.\\
Finally the massless Wess-Zumino model is treated with two independent couplings, the
Yukawa coupling and the coupling of the quartic interaction of the scalar and the
pseudoscalar field. One solution of the reduction equation corresponds to the
supersymmetric case considered by Wess and Zumino. In addition one finds a family of
solutions with an arbitrary parameter -- an exceptional case with an infinite number
of reduction solutions. A corresponding symmetry is not known.\\



\pagebreak

\noindent
\subsection{Relation between effective couplings for asymptotically free models}
Title: Relation between effective couplings for asymptotically free models\\
Authors:  R.\ Oehme, W.\ Zimmermann\\
Journal: \href{http://link.springer.com/article/10.1007/BF01221218}{Commun.\ Math.\ Phys.\ {\bf 97} (1985) 569-582}\\

Comment ({\sl Wolfhart Zimmermann}\,)  \\
Massless models of quantum field theory involving two couplings $g$ and $\lambda$
are considered which are renormalizable and asymptotically free. Momentum dependent
effective couplings $\bar{g}$ and $\bar{\lambda}$ (also called running coupling
parameters) are introduced by appropriate vertex functions at suitably chosen
momentum configurations. By the principle of asymptotic freedom the effective couplings
vanish in the high momentum limit. The purpose of this paper is to derive relations
between the effective couplings which aymptotically hold for large momenta or small
coupling values.\\
The momentum dependence of the effective couplings is controlled by the evolution
equations which are ordinary differential equations whith respect to the momentum
variable. By eliminating the momentum variable one obtains an ordinary differential
equation for $\bar{\lambda}$ as a function of $\bar{g}$ which has the form of a reduction
equation with the corresponding $\beta$-functions as coefficients. For studying the
high momentum behavior the $\beta$-functions are expanded with respect to powers of
$\bar{g}$ and $\bar{\lambda}$. It is assumed that powers of $\bar{\lambda}$ only are
absent in the expansion of the $\beta$-function associated with the coupling $\bar{g}$.
This should cover most applications. With the $\beta$-functions approximated to lowest
order the differential equation $\bar{\lambda}(\bar{g})$ can be solved exactly. Including
all higher powers one finds asymptotic expansions for $\bar{\lambda}(\bar{g})$ involving
powers (including fractional or irrational exponents) and possibly logarithmic terms.
The solutions obtained are complete in the sense as they generalize the exact solutions
found in lowest order.\\

\pagebreak

\subsection{Renormalization group equations with vanishing lowest order of the primary $\beta$-function}
Title: Renormalization group equations with vanishing lowest order of the primary \linebreak[4]
\hbox{\hspace{12mm}} $\beta$-function\\
Authors: R.\ Oehme, K.\ Sibold, W.\ Zimmermann\\
Journal: \href{http://ac.els-cdn.com/037026938490604X/1-s2.0-037026938490604X-main.pdf?_tid=beb7a23c-2b2c-11e3-840c-00000aacb361&acdnat=1380695873_9f72816c7293441a3d9a1bc4a6d4c981}{Phys.\ Letts.\ {\bf B147} (1984)115-120}\\

\noindent
Comment ({\sl Klaus Sibold}\,)\\
\noindent
Whereas in subsections 2.1, 2.2 the general method of reduction of couplings
has been exposed, in the present paper a class of theories is envisaged
which represents a special case only, but nevertheless is of quite some importance
for all applications to follow: it is assumed that the lowest order of
the primary $\beta$-function vanishes. This is of interest in supersymmetric
theories in particular. The study has been performed in massless models
with two couplings and it follows the pattern which had been suggested 
by QCD: the primary coupling is asymptotically free and one supposes
a secondary coupling to be given whose behavior is investigated as
dictated by its $\beta$-function. Here it only assumed that not all
coefficients of sixth order in the primary $\beta$-function vanish.
Then asymptotic behavior and stability of the solutions of the evolution
equations are derived.\\
The asymptotic behavior is studied under the assumption that the secondary
coupling considered as a function of the primary vanishes when the
primary tends to zero. As one of the results for supersymmetric Yang-Mills
theories with one Yukawa coupling constant for the interaction of chiral
superfields it turns out that they are unstable if they are UV-asymptotically
free. Here, as said above, the conclusion holds for the embedding into
a theory with two couplings.\\

\pagebreak
\subsection{Construction of gauge theories with a single coupling parameter for Yang-Mills and matter fields}
Title: Construction of gauge theories with a single coupling parameter for Yang-Mills and
\linebreak[4]\hbox{\hspace{12mm}}matter fields\\
Authors: R.\ Oehme, K.\ Sibold, W.\ Zimmermann\\
Journal: \href{http://ac.els-cdn.com/0370269385914169/1-s2.0-0370269385914169-main.pdf?_tid=b8d34952-2b36-11e3-a8ba-00000aacb35e&acdnat=1380700158_f9b9c7a40852e8290e56e8b600100200}{Phys.\ Letts.\ {\bf B153} (1985)142-146}\\

\noindent
Comment ({\sl Klaus Sibold}\,)\\ 
This paper continues via two examples the application of the reduction method
to construct in a neighbourhood of four couplings a gauge theory depending on
one coupling, the gauge coupling, only. The respective solutions of the 
reduction equations are power series in the remaining coupling, hence
strictly renormalizable.\\
The matter field content is chosen such that one of the examples can lead
to $N=2$ supersymmetry in a component formulation, the other one
to $N=4$ supersymmetry. And, indeed the respective values of the matter
couplings appear as solutions, hence to all orders of perturbation theory
there exist Green functions which depend on one coupling only and whose tree
approximation has the respective symmetry. Of  course nothing can be derived from
this analysis alone, on how the symmetry is realized in higher orders.\\
In both cases there exists a second solution, also to all orders, which does
not show supersymmetry. All of these solutions go to zero with the primary coupling. \\
A stability analysis along the lines of Lyapunov's theory has been performed.
The $N=2$ example is UV unstable. For the $N=4$ theory the system is UV-unstable
if $\beta \le 0$ and it is IR-unstable if $\beta \ge 0$ for small coupling.
Even after the proof that perturbatively the $\beta$-function vanishes
identically (cf.\ subsection 4.2) one cannot exclude terms which vanish
exponentially, hence the unequality assumptions are relevant.\\


\pagebreak
\subsection{Additional Remarks to Section 2}

\noindent
{\sl Klaus Sibold}\\
We first mention the review papers \cite{Oehme_tpsuppl}, \cite{Sibold_apprev} where
many examples and some
general discussion of the method have been presented.\\
Next we draw the readers attention to papers \cite{Kraus_abBRS}, \cite{Kraus_nonabBRS}, 
\cite{Kraus_beta}. As contribution to a systematic application of the reduction principle
they pose and answer the question how gauge theories live in a gauge-nonivariant 
surrounding. The free theory can, of course be analyzed and understood as consisting
of a gauge invariant and gauge fixing part leading to the well-known factor space structure
of the physical Hilbert space. When tackling the interacting theory by reduction of
gauge-noninvariant couplings non-linear gauge fixing has to be singled out which indeed
can be achieved as suggested by the gauge fixing parameter dependence of the free theory.
The abelian case can be mastered in full generality, whereas the non-abelian one
requires some additional assumption, either on the gauge fixing parameters or on the
complete model. E.g.\ demanding rigid gauge invariance suffices in the important
example of $SU(N)$ to find as unique solution of the
reduction equations the BRS invariant gauge theory with one coupling and a $\beta$-function
which is gauge parameter independent.\\
Most interesting is the result of the stability analysis (following Lyapunov's theory).
The eigenvalues of the stability matrix around the BRS-symmetric solution are complex
and change their (UV-, IR-) behavior depending on the value of the gauge fixing parameter.
Together with the results for the other examples examined in the present section the
following pattern for eigenvalues and general solutions arises:\\
\vspace{-5mm}
\begin{itemize}
\item for gauge theories: 
BRS-invariant theory embedded in non-invariant surrounding: eigenvalues complex.\\
Supersymmetric gauge theory embedded in non-supersymmetric surrounding: eigenvalues real;
general solutions exist which are not supersymmetric but still are power series with
integer exponents of the primary coupling. Asymptotic behavior fixed.\\
SYM with vanishing first order $\beta$-function of the gauge coupling, embedded
in non-supersymmetric surrounding: eigenvalues real; general solutions exist which
vanish exponentially for small coupling.\\
\item Models with spin $0,1/2$ only:
Field content not compatible with $N=1$ supersymmetry: eigenvalues real; general solution
with irrational exponents of the primary coupling.\\
Field content compatible with $N=1$ supersymmetry: eigenvalues real; general solutions 
with power series of integer powers of the primary coupling.\\
\end{itemize}
These regularities have not yet found any deeper understanding. In any case they underline
that for characterizing a specific solution of the renormalization group equations
one may either demand a symmetry or a power series in a primary coupling. One may very
rarely rely on an ``automatic'' realization via renormalization group flow.
This fact supports constructions of asymptotically vanishing solutions by
``partial reduction'' as used below in the standard model (subsection 3.3) and in its minimally
supersymmetric extension (subsection 5.10) .\\
 
\pagebreak

\noindent
\section{Reduction of couplings in the standard model}
\noindent
Comment ({\sl Klaus Sibold}\,)\\  
The following remarks form a general introduction to the above section\\
Even today, almost twenty years after our first paper on reduction of couplings in the standard
model the original motivation for applying this method to this model has not become obsolete,
neither by time nor by new insight.
The theoretical predictions originating from the standard model are in extremely good agreement
with experiment. Actually the most precisely measured physical quantities, the anomalous
magnetic moment of electron and myon agree within 3 parts per $10^{-9}$ with their prediction
by theory. Two decades of precision measurement and precision calculation yielded essentially
on all available observables a truely astonishing coincidence \cite{Hollikyellowrep}. And,
yet there is no convincing explanation  why the number of families is three; why the mass scales
-- the Planck mass and the electroweak breaking scale -- differ so much in magnitude, why the
Higgs mass is small compared with the Planck scale. And, quite generally, there is also no
explanation for the mixing of the families. \\
Reduction of couplings offers a way to understand at least to some degree masses and mixings
of charged leptons and quarks and the mass of the Higgs particle. It extends the well known
case of closed renormalization orbits due to symmetry to other, more general ones. Which
structure these orbits have had to be learned, i.e. deduced from the relevant renormalization
group equation in the specific model. In particular, one had to take into account the different
behavior of abelian versus non-abelian gauge groups and of the Higgs self-coupling,
say in the ultraviolet region. If asymptotic expansions should make sense in the transition
from a non-perturbative theory to a perturbative version it should be possible to rely on common 
ultraviolet asymptotic freedom. One also has to respect gross features coming from phenomenology.
In mathematical terms this is the problem of integrating
partial differential equations by imposing suitable boundary conditions (originating from physical
requirements): partial reduction.\\
And, indeed this is how we proceeded historically. In subsection 3.1 mixing of families
has been neglected and the structure in the space of running gauge, Higgs and Yukawa
couplings has been found, when asking for common ultraviolet behavior.
In subsection 3.2 quark family mixing has been analyzed, in subsection
3.3 the method of partial reduction has been introduced. (Actually, in subsection 5.2
this concept has been extended to couplings carrying dimensions.) In subsection 3.4 as an other,
additional ingredient we imposed the condition that quadratical divergencies be
absent. This requirement makes sense in the context of the standard model, because these
divergencies refer to a gauge invariant quantity. Remarkably enough, it turned out that
this postulate is indeed consistent with reduction.
Subsection 3.5 concludes these earliest investigations in the standard model with an update 
as of 1991. It yields as values for Higgs and top mass roughly 65, respectively 100 GeV.\\
Perhaps the most important and not obvious result of the entire analysis is the fact that reduction
of couplings (even the version of ``partial reduction'') is extremely sensitive to the model.
If one accepts the integration ``paths'' as derived in the papers of this section the
ordinary standard model can neither afford a mass of the top quark nor of the Higgs
particle as large as they have been found experimentally. The mismatch of the fact that
the experimental findings are in very good agreement with calculations and the fact that
the reduction paths of integration rule out the SM is only apparent: renormalization group
improvement of the theoretical predictions concerns essentially the QCD sector, where it
is taken into account in the reduction. Whereas the differences originating from the other couplings
turn out to be negligibly small.\\
Hence it became clear that other model classes are to be studied and further constraining
principles had to be found. This will be the subject of sections four and five.\\


\noindent
These earliest papers on reduction of couplings have been reviewed e.g.\ in  \\
\cite{Oehme_tpsuppl}, \cite{Sibold_apprev}.\\

\subsection{Higgs and top mass from reduction of couplings}
Title: Higgs and top mass from reduction of couplings\\
Authors: J.\ Kubo, K.\ Sibold, W.\ Zimmermann\\
Journal: \href{http://ac.els-cdn.com/055032138590639X/1-s2.0-055032138590639X-main.pdf?_tid=ce9a1062-2b37-11e3-add5-00000aacb35f&acdnat=1380700624_7629d3b3d5fc01273a0377f5521b9541}{Nucl.\ Phys.\ {\bf B259} (1985) 331-350}\\

\noindent
Comment ({\sl Klaus Sibold}\,)\\
In the context of the standard model with one Higgs doublet and $n$ families
the principle of reduction of couplings is applied. For simplicity mixing of
the families is assumed to be absent: the Yukawa couplings are diagonal and real.
For the massless model reduction solutions can be found to all orders of
perturbation theory as power series in the ``primary'' coupling, thus
superseding fixed point considerations based
on one-loop approximations. Due to the different asymptotic behaviour of the
$SU(3),SU(2)$ and $U(1)$ couplings the space of solutions is clearly structured
and permits reduction in very distinct ways only. Since reducing the gauge
couplings relative to each other is either inconsistent or phenomenologically
not acceptable, $\alpha_S$ (the largest coupling) has been chosen as the expansion 
parameter -- the primary coupling -- and thus UV-asymptotic freedom as the relevant
regime. This allows to neglect in the lowest order approximation the other
gauge couplings and to take their effect into account as corrections.\\
In the matter sector (leptons, quarks, Higgs) discrete solutions emerge for
the reduced couplings which permit essentially only the Higgs self-coupling
and the Yukawa coupling to the top quark to be non-vanishing.\\ 
Stability considerations (Liapunov's theory) show how the power series solutions
are embedded in the set of the general solutions. The free parameters in the
general solution represent the the integration constants over which one had
disposed in the power series, i.e.\ perturbative reduction solution. \\
Couplings of the massless model are converted into masses in the tree approximation
of the spontaneously broken model. For three generations one finds $m_H = 61$ GeV,
$m_{top} = 81$ GeV with an error of about 10-15\%.\\


\pagebreak
\noindent
\subsection{Quark family mixing and reduction of couplings}
Title: Quark family mixing and reduction of couplings\\
Authors: K.\ Sibold, W.\ Zimmermann\\
Journal: \href{http://ac.els-cdn.com/0370269387906344/1-s2.0-0370269387906344-main.pdf?_tid=62e211a2-2a9d-11e3-a82b-00000aacb35f&acdnat=1380634301_70a982e57037db9ff0abe096263ce436}{Phys.\ Letts.\ {\bf B191} (1987) 427-430}\\

\noindent 
Comment ({\sl Klaus Sibold}\,)\\
After having laid the groundwork for reduction in the standard model
in the paper of the previous subsection we continue this analysis 
by admitting the full-fledged Yukawa coupling matrices. In the case which
has been treated three families are being considered hence there appears
a complex $3\times3$ matrix $G^d$ for the down quarks and a similar matrix $G^u$
for the up quarks. Together with the Higgs coupling $\lambda$ they
are understood as functions of $\alpha_s$ which is the primary coupling following
the results of the previous paper. Hence we search for solutions of the
reduction equations which go to zero with $\alpha_s$, i.e.\ we impose
asymptotic freedom in the UV region.\\
The diagonal solutions of the non-trivial reduction which implied
non-vanishing masses for the top quark and the Higgs clearly also
govern the solution pattern for the mixing. For the trivial reduction
case arbitrary masses for the charged leptons and the quarks are
permitted. (Neutrinos are by assumption massless.) For the non-trivial
reduction, where the Higgs and top quark masses
are determined it is found that the Cabibbo angle is arbitrary,
mixing between the third and the first two families is however excluded.
This result is interesting indeed because the observed parameters in
the Kobayashi-Maskawa matrix which express mixing between the third and
the first two families are very small.\\
(Warning: The second equation of (6) in the paper contains a misprint.
The formula should read $c_{-}\not=0$.)



\pagebreak
\noindent
\subsection{New results in the  reduction of the standard model}
Title: New results in the  reduction of the standard model\\
Authors: J.\ Kubo, K.\ Sibold, W.\ Zimmermann\\
Journal: \href{http://ac.els-cdn.com/0370269389900348/1-s2.0-0370269389900348-main.pdf?_tid=eba308ac-2a9d-11e3-b24d-00000aab0f02&acdnat=1380634530_5ac1445197877a9c65da6a4d2d33853a}{Phys.\ Letts.\ {\bf B220} (1988) 185-191}\\

\noindent
Comment ({\sl Klaus Sibold}\,)\\
Reduction of couplings is based on the requirement that all reduced couplings vanish
simultaneously with the reducing -- the primary -- coupling. This is clearly only possible
if the couplings considered have the same asymptotic behavior or have vanishing $\beta$-functions.
Hence in the standard model, based on $SU(3)\times SU(2)\times U(1)$ straightforward reduction
cannot be realized. Since however the strong coupling $\alpha_s$ is, say at the $W$-mass, considerably larger
than the weak and electromagnetic coupling one may put those equal to zero, reduce within the system
of quantum chromodynamics including the Higgs and the Yukawa couplings and subsequently take
into account electroweak corrections as a kind of perturbation. This is called ``partial reduction''.
In the present paper a new perturbation method has been developed and then applied with the updated
experimental values of the strong coupling and the Weinberg angle.\\
If $\beta$ functions are non-vanishing they usually go to zero with some power of the couplings involved.
Thus, reduction equations are singular for vanishing coupling and require a case by case study
at this singular point. In particular this is true for the reduction equations of Yukawa and Higgs couplings
when reducing to $\alpha_s$. It is shown in the paper that for the non-trivial reduction solution
(i.e.\ only the top Yukawa coupling and the Higgs coupling do not vanish) one can de-singularize 
the system by a variable transformation and thereafter go over to a partial differential equation
which is easier to solve than the ordinary differential equations one started with. The reduction
solutions of the perturbed system are then in one-to-one correspondence with the unperturbed one's.\\
In terms of mass values the non-trivial reduction yields $m_t= 91.3$ GeV, $m_H=64.3$ GeV. These mass
values are at the same time the upper bound for the trivial reduction, where the Higgs mass is a function
of the top mass. Here we used as definition for ``trivial'' that the ratios of  top-Yukawa coupling,
respectively Higgs coupling to $\alpha_s$ go to zero for the weak coupling limit $\alpha_s$ going to zero.\\


\pagebreak
\noindent
\subsection{Cancellation of divergencies and reduction of couplings}
Title: Cancellation of divergencies and reduction of couplings
in the standard model\\
Authors: J.\ Kubo, K.\ Sibold, W.\ Zimmermann\\
Journal: \href{http://ac.els-cdn.com/037026938990035X/1-s2.0-037026938990035X-main.pdf?_tid=48eac4d2-2a9e-11e3-86da-00000aacb360&acdnat=1380634686_039c3c9d5ae52080b49b865f3560dd58}{Phys.\ Letts.\ {\bf B220} (1989) 191-194}\\

\noindent
Comment ({\sl Klaus Sibold}\,)\\
Although the standard model describes the experimental situation very well
it has (at least) two shortcomings which raise doubts that it can be considered
as a fundamental theory as opposed to an effective one. First, due to the quadratical
divergencies in the Higgs self-mass there is the problem of ``naturalness'',
also called hierarchy problem. Second, the masses of quarks and leptons as well
as the mixing angles enter as free parameters which have to be taken from experiment
-- these are unaesthetically many.\\
Reduction of couplings as described in the previous subsections indeed constrains the
parameters of the model. In the present paper it has been analyzed whether it is
possible to require in addition the absence of quadratical divergencies. If so, 
then the version with three families would indeed become strengthened as to be fundamental.\\
In order to proceed it has been shown first that postulating absence of quadratical
divergencies is a gauge and renormalization group invariant statement.
And, indeed the resulting constraint is compatible with reduction, at least with the
trivial one. The non-trivial reduction solution is however off by the uncertainties
of the measurement of $\alpha_{em}/\alpha_s$ and $sin^2\theta_W$.\\
Below, in section 5, the absence of quadratical divergencies will be implemented by
relying on supersymmetry and/or by soft breaking of susy which maintains their
absence. Hence this requirement and its interplay with reduction of couplings
remained substantial.\\


\pagebreak
\noindent
\subsection{ Precise determination of the top quark and Higgs masses}
Title: Precise determination of the top quark and Higgs masses
in the reduced standard \linebreak[4] \hbox{\hspace{12mm}} theory for electroweak and strong interactions\\
Author: J.\ Kubo\\
Journal: \href{http://ac.els-cdn.com/037026939190625Z/1-s2.0-037026939190625Z-main.pdf?_tid=9af508be-2a9e-11e3-9cf6-00000aab0f01&acdnat=1380634824_1ad1491b0629c8e99915c6dde4f2a4e2}{Phys.\ Letts.\ {\bf B262} (1991) 472-476}\\

\noindent
Comment ({\sl Jisuke Kubo}\,)\\
The top quark and Higgs mass, $
m_{t}$ and $m_{h}$, can be predicted within the standard model (SM)
when reduction of coupling constants (s.\ subsection 2.1) 
is applied.
At the one-loop order we obtained (s.\ subsection 3.1) 
 \begin{eqnarray*}
 m_{t} &\simeq& 81~\mbox{GeV}\ ,
 m_{h} \simeq 61~\mbox{GeV}~. 
 \end{eqnarray*}
 There are corrections to these values:
 \begin{enumerate}
 \item
 The above mass values depend on the
 SM parameters, in particular the strong coupling constant
  $\alpha_3$ and $\sin\theta_W$. Since the values of
 $\alpha_3$ and $\sin\theta_W$ have been updated,
 the above predictions   
 need to be updated, too.
 \item
 Two-loop corrections may be important.
 \item
 In subsection 3.1 
 the difference of the physical mass (pole mass)
 and the mass defined in the $\overline{\mbox{MS}}$ scheme has been ignored.
 \end{enumerate}

 In the present article all these corrections are included.
 We find that the correction coming from
 the $\overline{\mbox{MS}}$ to the pole mass transition
 increases $m_{t}$ by about 4 $\%$, while
 $m_{h}$ is increased by about 1 $\%$.
 The two-loop effect is non-negligible especially for $m_{t}$:
 +2  $\%$ for  $m_{t}$ and 0.2 $\%$ for $ m_{h}$.
 Taking into account all these corrections we obtain
 \begin{eqnarray*}
    m_{t} &=& 98.6\pm9.2~\mbox{GeV},
    m_{h} = 64.5\pm1.5 ~\mbox{GeV},  
 \end{eqnarray*}
 where  the  1991 values of $M_Z, \alpha_{3}(M_{Z}),
 sin^{2}\theta_{W}(M_{Z})$ and $\alpha_{em}(M_{Z})$
 are used.\\

 \noindent
 If we use their 2013 values given in \cite{Beringer:1900zz},
 we find that the change of the prediction is negligible.
 Obviously, this prediction   
 is inconsistent with the experimental
 observations. This may be seen as a good news, because
 we know that the SM has to be extended to explain
 the recent experimental observations such as the non-zero neutrino mass.
 Even a simplest extension to include a  dark matter candidate
 will change the 1991-prediction (which coincides essentially with a
 2013-prediction).\\



\pagebreak
\section{Abstract interludium}
Comment ({\sl Klaus Sibold}\,)\\
In the third section we presented the principle of reduction
of couplings and its application to the standard model.
These investigations took place, roughly, during the years 1983
until 1991. In parallel to them a program of renormalizing
supersymmetric theories was carried out which culminated
for models with one supersymmetric generator, $N=1$ in short,
in a fairly complete understanding of its maximal symmetry
content: superconformal symmetry. It turned out that in
all $N=1$ models the anomalies of the superconformal tranformations
lie in some susy multiplet and are provided by the supercurrent
and its moments in superspace. Next, it is crucial that a specific
$U(1)$ axial transformation, called $R$, forms part of the superconformal
algebra. For, axial transformations may lead to non-renormalization
theorems, which then affect the (non-)renormalization behavior of the 
anomalies of the other transformations.\\ 
\noindent
In the usual setup of perturbative quantum field theories
ultraviolet divergencies occur and have to be taken care of in
such a way that the fundamental postulates -- Lorentz
covariance, unitarity and causality -- are not violated. In
supersymmetric theories, as a rule, fewer divergencies show
up than in ordinary models of spin zero, one-half and one.
The non-abelian gauge theory with $N=4$ supersymmetries
has only one coupling, the gauge coupling. Its respective 
$\beta-$function automatically vanishes; this theory has been 
called ``finite``. In the more general case of $N=1$
supersymmetry one can now search if this can take place
by reducing the matter couplings 
to the gauge coupling, follow the effect of reduction
and combining the result with relations provided by the superconformal
symmetry. The non-renormalization theorems of axial current
anomalies yield then very interesting results. This refers
to subsections 4.1 and 4.2. (A somewhat non-technical report
on the outcome of these investigations is provided by \cite{Piguet_alushta}.)\\
In section 5 models will be considered which are based on 
supersymmetry and finiteness, i.e.\ the proliferation of free parameters
introduced by ``supersymmetrizing'' a phenomenologically viable
theory, say in order to suppress naturally quadratical divergencies,
is counterbalanced by restricting matter couplings via reduction
and asking for finiteness in the sense of having vanishing
$\beta$-functions. This application justifies the inclusion
of the respective papers in the present section.\\

\noindent
In subsection 4.3 a first step has been made towards incorporating
masses and gauge parameters when performing reduction of couplings:
it is shown that reduction of dimensionless couplings is possible
in the presence of such parameters.\\
These considerations are extended in subsection 4.4 to refer to the
notion of reduction itself by formulating the method also for ``couplings''
carrying dimension; this includes mass parameters. These investigations
provide the basis for the exploration and application of soft susy
breaking in the papers presented in section 5. Obviously nature is
not supersymmetric, but mechanisms for breaking supersymmetry are
rare. Dynamical mass generation is not easy to implement, spontaneous
breaking of susy does not lead very far, hence soft breaking which
maintains the benefits of susy is the most suitable tool. In
practice it has been found (s.\ section 5) that there exist also on the
level of soft terms closed renormalization orbits. Those can be
systematically searched for by reduction. It is then a matter of
detailed analysis to relate
(running) mass parameters to physical masses and to clarify the
different renormalization effects. Most important is the identification
of renormalization scheme independent quantities and resulting
calculational rules. \\

\subsection{Vanishing $\beta-$functions in $N=1$ supersymmetric 
gauge theories}
Title: Vanishing $\beta-$functions in supersymmetric gauge theories\\
Authors: Lucchesi, O. Piguet, K. Sibold\\
Journal: \href{http://retro.seals.ch/digbib/view?pid=hpa-001:1988:61::323}{Helv.\ Physica Acta {\bf 61} (1988) 321-344}\\

\noindent
Comment ({\sl Olivier Piguet}\,)\\
This paper presents a non-renormalization theorem for the vanishing, at
all orders of perturbation theory, of the Callan-Symanzik $\beta$-functions
for a class of $N=1$ supersymmetric non-abelian gauge theories where the gauge group
is simple. The matter content of the theory is assumed to be such that the anomaly
in the Slavnov-Taylor identity is absent, hence the gauge theory is consistent.
The necessary and sufficient conditions for the theorem to
hold are:\\
(i) the $\beta$-function of the gauge coupling vanishes in one-loop order;\\
(ii) the anomalous dimensions of the matter superfields vanish in one-loop order;\\
(iii) the Yukawa couplings of the matter supermultiplets solve as power series 
in the gauge coupling the Oehme-Zimmermann reduction equations (see Section 1).\\
The proof exploits
the supersymmetric correspondence of the conformal anomaly with a certain
axial current anomaly through the supercurrent multiplet. The theorem allows
the formulation of a simple criterion, involving only one-loop order quantities.
The outcome is a class of $N=1$ supersymmetric theories with a single coupling
constant which are “finite”, i.e., whose $\beta$-function vanish to all orders
of perturbation theory. An example based on the unitary group $SU(6)$ is
worked out, showing that this class of finite theories is not empty and contains
theories without extended supersymmetry.\\
\pagebreak
\subsection{Necessary and sufficient conditions for all order
vanishing $\beta-$functions in supersymmetric Yang-Mills
theories}
Title: Necessary and sufficient conditions for all order
vanishing $\beta-$functions in super-\linebreak[4]\hbox{\hspace{12mm}}symmetric Yang-Mills
theories\\
Authors: C.\ Lucchesi,\ O. Piguet, K.\ Sibold\\
Journal: \href{http://ac.els-cdn.com/0370269388902213/1-s2.0-0370269388902213-main.pdf?_tid=86aeb86c-2b31-11e3-bd60-00000aacb362&acdnat=1380697926_3e6e44f9c9c4d91e930e56a65854a2a9}{Phys.\ Letts.\ {\bf B201} (1988) 241-244}\\

\noindent
Comment ({\sl Klaus Sibold}\,)\\
Based on the theorems of the preceding subsection one-loop criteria
are given which are necessary and sufficient for the vanishing
of $\beta$-functions to all orders of perturbation theory.
They are operative in the fairly general setting of consistent $N=1$ supersymmetric
Yang-Mills theories. The following three conditions have to be satisfied:\\
(i) the $\beta$-function of the gauge coupling vanishes in one-loop order;\\
(ii) the anomalous dimensions of the matter superfields vanish in the one-loop order;\\
(iii) the Yukawa couplings solve the reduction equations (and satisfy (ii)) in such
a way that the solution is isolated and non-degenerate.\\
Isolation and non-degeneracy can usually be established (if not automatically true)
by imposing additional chiral symmetries or fixing arbitrary phases by hand: the
non-renormalization theorem for chiral vertices guarantees that they are not affected
by higher orders.\\
The second -- physicswise very interesting -- result of this paper is that it contains
an interpretation of what ``finiteness'' means. Vanishing $\beta$-functions say, of course,
that dilatations and special (super-)conformal symmetry are unbroken. Clearly also
$R$-invariance is maintained. But all other chiral symmetries which act as
outer automorphisms on susy are also unbroken: that their one-loop anomaly 
coefficients
vanish guarantees the compatibility of the equations used in condition (ii).
Hence one has a model which is free of all possible anomalies: those related to geometry
and those related to internal symmetries.\\
In section 5 the preceding criteria will be extensively used for finding finite theories
which are phenomenologically acceptable.\\
Another immediate application is possible
in investigations of anomalies via local coupling (with or without supergravity
background). Based on calculations in components within SYM with local gauge coupling 
\cite{Kraus_susy_anom1}, \cite{Kraus_susy_anom2} an anomaly had been found
and attributed to supersymmetry. For a
manifestly supersymmetric gauge in the analogous study by \cite{KrausRuppSibold}
it was realized that this anomaly could be shifted into a renormalization of the
$\theta$-angle. Remarkably enough, in a finite SYM theory this anomaly is absent and thus the
$\theta$-angle is not renormalized.\\
It is then tempting to speculate that amongst such finite $N=1$ models there
is (at least) one which permits to cancel the Weyl anomaly in conformal supergravity
theory. That, in turn might permit to construct power counting renormalizable
theories containing quantized gravity. (As a guide to the rich literature one may consult
\cite{Tseytlin}.)\\


\pagebreak

%
%
%
\subsection{Reduction of couplings in the presence of parameters}
Title: Reduction of couplings in the presence of parameters\\
Authors: O.\ Piguet, K.\ Sibold\\
Journal: \href{http://ac.els-cdn.com/0370269389901603/1-s2.0-0370269389901603-main.pdf?_tid=f80a9f60-be36-11e3-8547-00000aacb361&acdnat=1396863085_e90ff456a167d77c5bf508159108935e'}{Phys.\ Letts.\ {\bf B 229} (1989) 83-89}\\ 

\noindent
Comment ({\sl Klaus Sibold, Wolfhart Zimmermann}\,)\\
In the papers on reduction and its application in the above sections two and three
reduction had been performed for massless theories. It is however obvious
that reduction is of considerable interest also in massive theories and
in particular reduction of couplings carrying dimension is a very important
issue (s.\ section 5). In the present paper this problem has been addressed in its simplest
version: in a gauge theory mass parameters $m_a$ and a gauge fixing
parameter $\alpha$
are permitted, where masses are fixed on-shell and matter couplings
are fixed by $\alpha$-independent normalization conditions. It is also
necessary to introduce a special value $\alpha_0$ for the gauge parameter
$\alpha$ in addition to the standard normalization point parameter
$\kappa$. \\
Due to the presence of mass parameters one has now to distinguish between
renormalization group and Callan-Symanzik equations. All $\beta$-functions
can be rendered $\alpha$-independent to all orders, independent of
$\alpha_0$ to one-loop order and the Callan-Symanzik $\beta$-functions
mass independent to one-loop. The $\beta$-functions of the renormalization
group equations will in general depend on mass ratios already in one loop.\\
When setting up reduction equations for the dimensionless coupling parameters
those for the renormalization group equations turn out to involve partial
derivatives with respect to mass values. But for the Callan-Symanzik equation,
fortunately, they take the form of ordinary differential equations quite similar
to the massless case with only parametric dependence on the mass and
gauge fixing values. The problem of consistency between potentially different
solutions originating from either renormalization group respectively Callan-Symanzik
equation  can be solved by employing the consistency
of the original differential equations referring to the original parameters:
one can show that the reduced couplings satisfy the required differential
equations (namely variations with respect to $\alpha, \alpha_0, \kappa$)
for power series solutions of the reduction equations. Hence these reduced
theories can be considered as renormalizable field theories. Furthermore the
mass dependence of the RG $\beta$-functions in order $n-1$ determines the
mass dependence of the reduction solution in order $n$. For more general
solutions this is unlikely to happen.\\
The general case will be presented in the next subsection.\\


\pagebreak


\noindent
\subsection{Scheme independence of the reduction principle and
asymptotic freedom in several couplings}
Title: Scheme independence of the reduction principle and asymptotic freedom in several
       \linebreak[4]\hbox{\hspace{12mm}}couplings \\
Author: W.\ Zimmermann\\
Journal: \href{http://link.springer.com/article/10.1007/s002200100396}{Commun.\ Math.\ Phys.\ {\bf 219} (2001) 221-245}\\


\noindent
Comment ({\sl Wolfhart Zimmermann}\,)\\
For renormalizable models of quantum field theory there is considerable arbitrariness
in setting up schemes of renormalization. But different schemes should be equivalent in
the sense that Green's functions -- apart from normalization factors of the fields -- become
identical after an appropriate transformation of the coupling parameters. For the
reduction principle to be a meaningful concept it must be invariant under such scheme
changing transformations. The freedom of choosing a convenient renormalization scheme
may be used to simplify the form of conditions for the reduction principle to hold.\\
In the first part of the present work the scheme independence of the reduction principle
is proved. Apart from dimensionless couplings, pole masses and gauge parameters the
model may also involve coupling parameters carrying a dimension and variable masses.
Pole masses refer to the lowest propagator singularities, variable masses are defined by
propagators at the normalization point and treated like couplings with dimension. Since
relevant for some applications also partial reductions are included. Accordingly, some of
the couplings are selected as primary couplings on which the remaining reduced couplings
depend. The reduction principle states that Green's functions expressed in terms of the
primary couplings satisfy the corresponding renormalization group equations. In addition,
it is required that all couplings simultaneously vanish in the weak coupling limit and
allow for power series expansions in the primary couplings. All these requirements are
shown to be invariant under scheme changing transformations thus establishing the scheme
independence of the reduction principle.\\
As an application massive models of quantum field theory are treated with several
dimensionless couplings. One of them is selected as primary coupling on which the other
couplings depend according to the reduction principle. A transformation of the coupling
parameters is constructed for defining an equivalent renormalization scheme in which the
original $\beta$-functions are replaced by their massless limits. Due to the scheme
independence the reductions equations also hold in the new renormalization scheme with
mass independent $\beta$-functions as coefficients. Their final form is a set of ordinary 
differential equations with only parametric dependence on the masses.\\

\noindent
The last part of this work concerns the property of asymptotic freedom for models
involving several couplings. Renormalizable models of quantum field theory are studied
with positive dimensionless coupling parameters. Effective couplings are introduced by
appropriate vertex functions. Their momentum dependence is controlled by the evolution
equations, a system of ordinary differential equations in the momentum variable with the
$\beta$-functions as coefficients. Asymptotic freedom states that all effective couplings 
simultaneously vanish in the high momentum limit. As a consequence all $\beta$-functions
are negative in the domain considered. For models with only one coupling the negative sign
of the $\beta$-function is also a sufficient condition for asymptotic freedom. In case of
several couplings asymptotic freedom is not a property of the model as such, but selects 
particular solutions of the system by placing constraints on the coupling parameters.
These are obtained by eliminating the momentum variable in the evolution equations. To this
end the momentum variable is replaced by one of the effective couplings, called the primary
coupling, as independent variable. With this substitution the evolution equations take the
form of reduction equations for the other effective couplings (the reduced couplings) as
functions of the primary coupling. The momentum dependence is then regulated by the remaining
evolution equation of the primary coupling with negative $\beta$-function. For asymptotic freedom
to hold the reduced couplings must vanish with the primary coupling in the weak coupling limit
(or high momentum limit) in accordance with the reduction principle.\\


\pagebreak


\newcommand{\be}{\begin{equation}}
\newcommand{\ee}{\end{equation}}
\newcommand{\brat}{{\rm BR}}
\newcommand{\bea}{\begin{eqnarray}}
\newcommand{\eea}{\end{eqnarray}}
\include{paperdef}

\newcommand{\tev}{\,\, \mathrm{TeV}}
\newcommand{\gev}{\,\, \mathrm{GeV}}
\newcommand{\mt}{m_t}
\newcommand{\mb}{m_b}
\newcommand{\br}{{\rm BR}}
\newcommand{\ga}{\gamma}
\newcommand{\si}{\sigma}
\newcommand{\Mh}{M_h}
\newcommand{\MH}{M_H}


\section{Phenomenologically viable models; finiteness; top and
	Higgs mass predictions agreeing with experiment}

Comment ({\sl Myriam Mondrag\'on, George Zoupanos}\,)\\
Let us first give a general introduction to this section.\\
In the recent years the theoretical endeavours that attempt to
achieve a deeper understanding of Nature have presented a series
of successes in developing frameworks such as String Theories and
Noncommutativity that aim to describe the fundamental theory at
the Planck scale. However, the essence of all theoretical efforts
in Elementary Particle Physics (EPP) is to understand the present
day free parameters of the Standard Model (SM) in terms of few
fundamental ones, i.e. to achieve {\it reductions of couplings}.
Unfortunately, despite the several successes in the above frameworks
they do not offer anything in the understanding of the free paramaters
of the SM.  The pathology of the plethora of free parameters is deeply
connected to the presence of  {\it infinities} at the quantum level.
The renormalization program can remove the infinities by introducing
counterterms, but only at the cost of leaving the corresponding terms as
free parameters. 
 To reduce the number of free parameters of a theory, and thus render it
 more predictive, one is usually led to introduce a symmetry. Grand Unified
 Theories (GUTs) are very good examples of such a procedure. For instance,
 in the case of minimal $SU(5)$, because of the (approximate) gauge coupling
 unification, it was possible to reduce the gauge couplings of the SM to one.
 In fact, the LEP data suggested that a further symmetry, namely $N = 1$ global
 supersymmetry should also be required to make the prediction viable. GUTs can
 also relate the Yukawa couplings among themselves, again $SU(5)$ provided an
 example of this by predicting the ratio $M_\tau/M_b$ in the SM. Unfortunately,
 requiring more gauge symmetry does not seem to help, since additional
 complications are introduced due to new degrees of freedom, in the ways and
 channels of breaking the symmetry, among others. Therefore, the fundamental
 lesson we have learned from the extensive studies of GUTs was that unification
 of gauge couplings is a very good idea, which moreover is nicely realized in
 the minimal supersymmetric version of the Standard Model (MSSM). In addition
 the use of the renormalization group equations (RGEs) has been established as
 the basic tool in the corresponding studies.\\
 A natural extension of the GUT idea is to find a way to relate the gauge and
 Yukawa sectors of a theory, that is to achieve gauge-Yukawa Unification (GYU)
 that will be presented in the subsections 5.1, 5.2, 5.5.  Following the original
 suggestion for reducing the couplings discussed in the previous sections, within
 the framework of GUTs we were hunting for renormalization group invariant (RGI) 
 relations holding below the Planck scale, which in turn are preserved down to the
 GUT scale.  It is indeed an impressive observation that one can guarantee the
 validity of the RGI relations to all-orders in perturbation theory by studying
 the uniqueness of the resulting relations at one-loop (sect.~2).  Even more remarkable
 is the fact that it is possible to find RGI relations among couplings that guarantee
 finiteness to all-orders in perturbation theory (sect.~3). The above principles have
 only been applied in $N = 1$ supersymmetric GUTs for reasons that will be transparent
 in the following subsections, here we should only note that the use of $N = 1$ supersymmetric
 GUTs comprises the demand of the cancellation of quadratic divergencies in the SM.  The above
 GYU program applied in the dimensionless couplings of supersymmetric GUTs had already
 a great success by predicting correctly, among others, the top quark mass in the finite
 $N = 1$ supersymmetric $SU(5)$ before its discovery \cite{Lancaster:2011wr}.
 
 Although supersymmetry seems to be an essential feature for a successful realization of
 the above program, its breaking has to be understood too, since it has the ambition to
 supply the SM with predictions for several of its free parameters. Indeed, the search for
 RGI relations has been extended to the soft supersymmetry breaking sector (SSB) of these
 theories, which involves
parameters of dimension one and two. In addition, there was important progress  concerning
the renormalization properties of the SSB parameters, based on the powerful supergraph method
for studying supersymmetric theories, and it was applied to the softly broken ones by using the
``spurion'' external space-time independent superfields. 
According to this method a softly broken supersymmetric gauge theory is considered as a
supersymmetric one in which the various parameters, such as couplings and masses, have
been promoted to external superfields. Then, relations among the soft term renormalization
and that of an unbroken supersymmetric theory have been derived. In particular the
$\beta$-functions of the parameters of the softly broken theory are expressed in terms
of partial differential operators involving the dimensionless parameters of the unbroken
theory. The key point in solving the set of coupled differential equations so as to be
able to express all parameters in a RGI way, was to transform the partial differential
operators involved to total derivative operators.  It is indeed possible to do this by
choosing a suitable RGI surface.

On the phenomenological side there exist some serious developments too.  Previously an
appealing ``universal'' set of soft scalar masses was assumed in the SSB sector of
supersymmetric theories, given that apart from economy and simplicity (1) they are
part of the constraints that preserve finiteness up to two-loops, (2) they appear
in the attractive dilaton dominated supersymmetry breaking superstring scenarios.
However, further studies have exhibited a number of problems, all due to the restrictive
nature of the ``universality'' assumption for the soft scalar masses. Therefore, there were
attempts to relax this constraint without loosing its attractive features. Indeed an
interesting observation on $N = 1$ GYU theories is that there exists a RGI sum rule for
the soft scalar masses at lower orders in perturbation theory, which was later extended to
all-orders, and that manages to overcome all the unpleasant phenomenological consequences.
 Armed with the above tools and results we were in a position to study the spectrum of the
 full finite models in terms of few free parameters, with emphasis on the predictions of
 supersymmetric particles and the lightest Higgs mass.  The result was indeed very
 impressive since it led to a prediction of the Higgs mass which coincided  with the
 results of the LHC for the Higgs mass, 
$125.5 \pm 0.2 \pm 0.6 \gev$ by ATLAS \cite{Aad:2012tfa+ATLAS:2013mma} and
$125.7 \pm 0.3 \pm 0.3 \gev$~by CMS \cite{Chatrchyan:2012ufa+CMS:yva}, 
and  predicted a supersymmetric spectrum  consistent with the non-observation of
coloured supersymmetric particles at the LHC. 
These successes will be presented in subsections 5.5, 5.8 and 5.9.

 Last but certainly not least, the above machinery has been recently applied in the
 MSSM with impressive results concerning the predictivity of
 the top, bottom and Higgs masses, being at the same time consistent with the
 non-observation of supersymmeric particles at the LHC. These results 
 will be presented in subsection 5.10.

\pagebreak
\noindent
\subsection{Finite unified models}
Title: Finite unified models\\
Authors: D.\ Kapetanakis, M.\ Mondragon, G.\ Zoupanos\\
Journal: \href{http://download.springer.com/static/pdf/553/art%253A10.1007%252FBF01650445.pdf?auth66=1400919343_252ef26ee7ff2e1bc3e6a6603b54eabb&ext=.pdf}{Z.\ Phys.\ {\bf C60} (1993) 181-186}\\ 
\noindent	
Comment ({\sl Myriam Mondrag\'on, George Zoupanos}\,)\\	
 The principle of finiteness  requires perhaps some more motivation
 to be considered and generally accepted these days than when it was
 first envisaged.
 It is however interesting to note that in the old days the general
 feeling was quite different. Probably the well
 known Dirac's phrase that ``...divergencies are hidden under the carpet'' is
 representative of the views of that time.
 In recent years we have a more relaxed attitude towards divergencies. Most
 theorists believe that the divergencies are signals of the existence of a higher
 scale, where new degrees of freedom are excited.
 Even accepting this dogma, we are naturally led to the conclusion that
 beyond the unification scale, i.e. when all interactions have been taken
 into account in a unified scheme, the theory should be completely finite.
 In fact, this is one of the main motivations and aims of string, non-commutative
 geometry, and quantum group theories, which include also gravity in the unification
 of the interactions.
 In our work on reduction of couplings and finiteness we restricted ourselves to 
 unifying only the known gauge interactions, based on a lesson of the history of 
 EPP that if a nice idea works in physics, usually it is realised in its simplest 
 form. Finiteness is based on the fact that it is possible to find renormalization
 group invariant (RGI) relations among couplings that keep finiteness in perturbation
 theory, even to all orders.
 Accepting finiteness as a guiding principle in constructing realistic theories of EPP,
 the first thing that comes to mind is to look for an $N = 4$ supersymmetric unified 
 gauge theory, since these theories are finite to all-orders for any gauge group. 
 However nobody has managed so far to produce realistic models in the framework of
 $N = 4$ SUSY. In the best of cases one could try to do a drastic truncation of the
 theory like the orbifold projection of refs.~\cite{Kachru:1998ys,Chatzistavrakidis:2010xi},
 but this is already a different theory than the original one. The next possibility is to
 consider an $N = 2$ supersymmetric gauge theory, whose beta-function receives corrections
 only at one-loop. Then it is not hard to select a spectrum to make the theory all-loop
 finite. However a serious obstacle in these theories is their mirror spectrum, which in
 the absence of a mechanism to make it heavy, does not permit the construction of realistic
 models. Therefore, we are naturally led to consider $N = 1$ supersymmetric gauge theories,
 which can be chiral and in principle realistic.\\
 Before our work the studies on $N = 1$ finite theories were following two directions:
 (a) construction of finite theories up to two-loops examining various possibilities to
 make them phenomenologically viable,
 (b) construction of all-loop finite models without particular emphasis on the
 phenomenological consequences. The success of our work was that we  constructed the
 first realistic all-loop finite model, based on the theorem presented in the subsection
 4.1, realising in this way an old theoretical dream of field theorists. Equally
 important was the correct prediction of the top quark mass one and half year before
 the experimental discovery. It was the combination of these two facts that motivated
 us to continue with the study of $N=1$ finite theories.  It is worth noting that nobody
 expected at the time such a heavy mass for the top quark. Given that the analysis of
 the experimental data changes over time, the comparison of our original prediction with
 the updated analyses will be discussed later, in particular in subsection 5.8.
 

\pagebreak


\subsection{Reduction of couplings and heavy top quark in the minimal SUSY GUT}
Title: Reduction of couplings and heavy top quark in the minimal SUSY GUT\\
Authors: J.\ Kubo, M.\ Mondragon, G.\ Zoupanos\\
Journal:  \href{http://ac.els-cdn.com/0550321394902968/1-s2.0-0550321394902968-main.pdf?_tid=4f09bd60-2b38-11e3-ae66-00000aab0f02&acdnat=1380700839_e8fdc28890f37e0950732cef211d7436}{Nucl.\ Phys.\ {\bf B424} (1994) 291-307}\\

\noindent
Comment ({\sl Myriam Mondrag\'on, George Zoupanos}\,) \\ 
To start with, it would have been natural to write this paper before the construction of
$N = 1$ Finite Unified Models which were discussed in the previous subsection. This work is very interesting for a number of reasons. 
The $N = 1$ minimal supersymmeric $SU(5)$ was logically the minimal framework to discuss  the reduction of coupling ideas in a realistic supersymmetric unification setup, the only known consistent framework
to overcome the problem of quadratic divergencies of the SM and also the first unification attempt.
Another interesting aspect of this study was to to examine to which extent the prediction ot the top quark mass of the Finite models was persisting in other GUTs as a more general feature of the reduction of couplings, which led to an exhaustive search for GYU in $N = 1$ supersymmetric GUTs. 
 Finally, the $N = 1$ minimal supersymmetric $SU(5)$ GUT is a nice framework to realize physically and apply technically the idea of {\it partial reduction} initiated in subsections 3.3 and 3.5. More specifically, in the study of Finite models a complete reduction of couplings was achieved, which was not expected to be the case in the minimal supersymmetric $SU(5)$. On the other hand the method of partial reduction, already introduced in subsection 3.1 became more transparent, especially after the reduction equations had been replaced by the mathematically equivalent set of partial differential equations as described in subsections 3.3 and 3.5. Therefore, the minimal supersymmetric $SU(5)$ was a natural new framework for an innovative method to be applied.
A rather interesting feature that emerged is that of all the possible solutions only two are asymptotically free, and both of them lie in the same RGI surface.  Even more remarkable is that they lead to good phenomenology, compatible with the data available at the time.

 In the future it is worth to have a fresh look to the reduction of couplings in the minimal $N = 1$ supersymmetric $SU(5)$, including the soft supersymmetry sector, in view of the results of the corresponding search in the MSSM to be discussed in subsection 5.10 and the updated experimental results on the top and bottom quark masses, as well as the discovery of the Higgs particle at LHC.


\pagebreak

\noindent
\subsection{Perturbative unification of soft supersymmetry-breaking}
Title: Perturbative unification of soft supersymmetry-breaking
terms\\
Authors: J.\ Kubo, M.\ Mondragon, G.\ Zoupanos\\
Journal: \href{http://ac.els-cdn.com/S0370269396013238/1-s2.0-S0370269396013238-main.pdf?_tid=073b33e0-2a9f-11e3-9c84-00000aab0f26&acdnat=1380635006_f5e5b9487acf4cb7c84ab1475dcd0ac3}{Phys.\ Letts.\ {\bf B389} (1996) 523-532}


\noindent
Comment ({\sl Myriam Mondrag\'on, George Zoupanos}\,)\\
As we have seen in subsections 2.1 and 2.2 the reduction of couplings was originally formulated for massless theories. On the other hand the successful reduction and impressive predictions of the top and bottom quark masses of $N = 1 ~SU(5)$ GUTs (finite and minimal supersymmetric) require the introduction of a massive soft supersymmery breaking (SSB) sector to become realistic.
 The extension of the reduction of couplings to theories with
massive parameters is not straightforward if one wants to keep the generality and the rigour on the same level as for the massless case. 
In this paper for simplicity a mass-independent renormalization scheme has been employed so that all the RG functions have only trivial dependencies on the dimensional parameters. Then the method suggested consists in searching for RGI relations among the SSB parameters, which are consistent with the perturbative renormalizability.

The method has been applied in the minimal GYU $N = 1$ supersymmetric $SU(5)$ model with the result that the SSB sector contains as the only arbitrary parameter the unified gaugino mass. Another characteristic feature of the findings of the analysis is that the set of the perturbatively unified SSB parameters differs significantly from the so-called universal SSB parameters, signaling already at that time the existence of a ``sum rule'' in GYU theories, as will be discussed later in subsections 5.5 and 5.6. The mass spectrum was then calculated using the experimental constraints known at the time and would have been ruled out now with the present LHC results. A new analysis, taking into account the recent B-physics results and including the radiative corrections coming from the supersymmetric spectrum for the bottom and tau masses, certainly would be very interesting and could lead to different spectrum to be compared with the recent findings at LHC on the Higgs mass and on the bounds of supersymmetric particles.


\pagebreak


\subsection{Unification beyond GUTs: Gauge Yukawa unification (Lectures)}
Title: Unification beyond GUTs: Gauge Yukawa unification\\
Authors: J.\ Kubo, M.\ Mondragon, G.\ Zoupanos\\ 
Journal: \href{http://www.actaphys.uj.edu.pl/_cur/store/vol27/pdf/v27p3911.pdf}{Acta\ Phys.\ Polon.\ {\bf B27} (1997) 3911-3944}\\


\noindent
Comment ({\sl Myriam Mondrag\'on, George Zoupanos}\,)\\
As has been already noted a natural extension of the GUT idea is to find a way to relate the gauge and Yukawa sectors of a theory, that is to achieve GYU. A symmetry which naturally relates the two sectors is
supersymmetry, in particular $N = 2$ supersymmetry. However, as has been also noted earlier in a different context, $N = 2$ supersymmetric theories have serious phenomenological problems due to light mirror fermions. Also in superstring theories and in composite models there exist relations among the gauge and Yukawa couplings, but both kind of theories have phenomenological problems, which we are not going to address here.

 There have been other attempts to relate the gauge and Yukawa
sectors which we recall and update for completeness here, while the references are already in the lectures paper.
One was proposed by Decker, Pestieau, and Veltman. By requiring the absence of quadratic divergencies in the SM,
they found a relationship among the squared masses appearing in the Yukawa and in the gauge sectors of the theory.
A very similar relation is obtained  by applying naively in the SM the general formula derived from demanding
spontaneous supersymmetry breaking via F-terms. In both cases a prediction for the top quark was possible only
when it was permitted experimentally to assume the $M_H  \ll M_{W,Z}$ with the result $M_t \simeq 69$ GeV.
Otherwise there is only a quadratic relation among $M_t$ and $M_H$. Using this relationship in the former case
and a version of naturalness into account, i.e. that the quadratic corrections to the Higgs mass be at most equal
to the physical mass, the Higgs mass is found to be
$\sim 260$ GeV, for a top quark mass of around 176 GeV, in complete disagreement with the recent findings at
LHC \cite{Aad:2012tfa+ATLAS:2013mma, Chatrchyan:2012ufa+CMS:yva}.

 A well known relation among gauge and Yukawa couplings is the
Pendleton-Ross (P-R) infrared fixed point. The P-R proposal, involving the Yukawa coupling of the top quark $g_t$ and the strong gauge coupling $\alpha_3$, was that the ratio $\alpha_t/\alpha_3$, where $\alpha_t = g_t^2/4\pi$, has an
infrared fixed point. 
This assumption predicted $M_t \sim 100$ GeV. In addition, it has been shown that the P-R conjecture is not justified at two-loops, since the ratio $\alpha_t/\alpha_3$ diverges in the infrared.
Another interesting conjecture, made by Hill, is that $\alpha_t$ itself
develops a quasi-infrared fixed point, leading to the prediction
$M_t \sim 280$ GeV.
 The P-R and Hill conjectures have been done in the framework of the SM.
The same conjectures within the Minimal Supersymmetric SM (MSSM) lead to the following relations:
$$
   M_t \approx 140 \gev \sin\beta \textrm{(P-R)},\quad M_t \approx 200 \gev \sin\beta \textrm{(Hill)},
$$
where $\tan\beta = v_u/v_d$ is the ratio of the two vacuum expectation values (vev's) of the Higgs fields of the MSSM. From theoretical considerations one can expect
$$
    1 < \tan\beta < 50  \Leftrightarrow 1/\sqrt{2} < \sin\beta < 1.
$$
This corresponds to
$$
 100 ~\gev < M_t < 140 \gev ~ \textrm{ (P-R)}, \quad 140 \gev < M_t < 200 \gev \textrm{ (Hill)}.
$$

Thus, the MSSM P-R conjecture is ruled out, while within the MSSM, the Hill conjecture does not give a prediction for $M_t$, since the value
of $\sin\beta$ is not fixed by other considerations. The Hill model can accommodate the correct value of $M_t \sim 173$ GeV for $\sin\beta \approx 0.865$
corresponding to $\tan\beta \approx 1.7$. Such small values, however, are strongly 
challenged if the newly discovered Higgs particle is identified with the lightest MSSM Higgs boson \cite{Heinemeyer:2011aa}.
Only a very heavy scalar top spectrum with large mixing could accommodate such a small $\tan\beta$ value.

The consequence of GYU is that in the lowest order in perturbation theory
the gauge and Yukawa couplings above $M_{GUT}$ are related in the form
\begin{equation*}
	g_i = \kappa_i g_{GUT}, \hspace{3mm}i = 1, 2, 3, e, ..., \tau, b, t, \hspace{5cm}(*)          
\end{equation*}
where $g_i$ (i = 1,...,t) stand for the gauge and Yukawa couplings, $g_{GUT}$ is
the unified coupling and we have neglected the Cabbibo-Kobayashi-Maskawa
mixing of the quarks. So,  eq.~$(*)$ 
 corresponds to a set of boundary conditions on the renormalization group evolution for the effective theory below $M_{GUT}$, which we have assumed to be the MSSM. As we have seen in subsections 5.1 and 5.2 it is possible to construct supersymmetric GUTs with GYU in the third generation that can predict the bottom and top quark masses in accordance with the experimental data. This means that the top-bottom hierarchy could be explained in these models, in a similar way as the hierarchy of the gauge couplings of the SM can be explained if one assumes the existence of a unifying gauge symmetry at $M_{GUT}$.
 It is clear that the GYU scenario is the most predictive scheme as far
as the mass of the top quark is concerned. It may be worth recalling the
predictions for $M_t$ of ordinary GUTs, in particular of supersymmetric
$SU(5)$ and $SO(10)$. The MSSM with $SU(5)$ Yukawa boundary unification allows
$M_t$ to be anywhere in the interval between 100-200 GeV for varying $\tan\beta$,
which is now a free parameter. Similarly, the MSSM with $SO(10)$ Yukawa
boundary conditions, i.e. $t-b -\tau$  Yukawa Unification, gives $M_t$
in the interval 160-200 GeV. In addition we have analyzed \cite{Kubo:1995cg} the infrared quasi-fixed-point behaviour of the $M_t$ prediction in some detail.
In particular we have seen that the infrared value for large $\tan\beta$ depends on $\tan\beta$ and its lowest value is $\sim 188$ GeV. Comparing this with the experimental
value $m_t= (173.2 \pm 0.9)$ GeV \cite{Lancaster:2011wr} we conclude that the present data on $M_t$ cannot be explained from the infrared quasi-fixed-point behaviour alone (see Figure 4 of hep-ph/9703289). An estimate of the theoretical uncertainties involved in GYU has been done in ref \cite{Kubo:1995cg}.  Although a fresh look has to be done in the case of the minimal $N = $1 supersymmetric $SU(5)$, we can conclude that the studies on the GYU of the asymptotically non-free supersymmetric Pati-Salam \cite{Kubo:1994xa} and asymptotically non-free $SO(10)$ \cite{Kubo:1995zg} models have ruled them out on the basis of the top quark mass prediction.

 It sould be emphasized once more that only one of the Finite Unified models (discussed in subsection 5.1 and which will be further discussed  in sections 5.5, 5.8, 5.9) not only predicted correctly the top and bottom quark masses but in addition predicted the Higgs mass in striking agreement with the recent findings at LHC \cite{Aad:2012tfa+ATLAS:2013mma,Chatrchyan:2012ufa+CMS:yva}.


\pagebreak


\noindent
\subsection{Constraints on finite soft supersymmetry-breaking terms}
Title: Constraints on finite soft supersymmetry-breaking terms.\\
Authors: T.\ Kobayashi, J.\ Kubo, M.\ Mondragon, G.\ Zoupanos\\
Journal: \href{http://ac.els-cdn.com/S0550321397007657/1-s2.0-S0550321397007657-main.pdf?_tid=29e6fbc2-2b3a-11e3-ae66-00000aab0f02&acdnat=1380701636_38234c175b52752d1808e4b8d2b3dd46}{Nucl.\ Phys.\ {\bf  B511} (1998) 45-68} 


\noindent
Comment ({\sl Myriam Mondrag\'on, George Zoupanos}\,)\\
 This is one of the most important and complete papers written on the subject of Finite Unified Theories and their predictions.
 An important point is that a new $N = 1$ Finite $SU(5)$ model was suggested, which (a) is more economical in the number of free parameters as compared to the original discussed in subsection 5.1 (it contains two instead of three parameters in its SSB), and (b) the new Finite model gives more accurate predictions for the top and bottom quark masses as seen today. At the time both Finite $SU(5)$ models were consistent with experimental data, but in a more recent analysis that will be presented in subsection 5.8 only a version of the second one survives in the comparison with the updated top and bottom quark mass measurements.

 Another important issue discussed in the present paper concerns the ``sum rule'' for the soft scalar masses at two loops.
To be more specific a number of problems appeared in finite unified theories using the attractive ``universal'' set of soft scalar masses. For instance, (i) the universality predicted that the lightest supersymmetric
particle was a charged particle, namely the superpartner of the $\tau$  lepton $\tilde{\tau}$, (ii) the MSSM with universal soft scalar masses was inconsistent with the standard radiative electroweak symmetry breaking, and (iii) which is the worst of all, the universal soft scalar masses lead to charge and/or colour breaking minima deeper than the standard vacuum. Naturally there have been attempts to relax this constraint. First an interesting observation was made that in a general $N = 1$ gauge-Yukawa unified (GYU) theories there exists a RGI ``sum rule'' for the soft scalar masses at one-loop, which obviously holds for the finite theories too. In the present paper it was found that in finite theories the ``sum rule'' remains RGI at two-loops with the surprising result that it does not change from its one-loop form, i.e. it does not receive two-loop corrections. In addition, some interesting remarks were done concerning the relation of the sum rule to certain string models.

 Eventually in the present paper it was presented a complete analysis of the two Finite Unifite $SU(5)$ theories and their phenomenological consequences.
The MSSM with the finiteness boundary conditions at the unification scale was examined by studying the evolution of the dimensionless parameters at two loops and the dimensionful at one loop. As a result it was determined the parameter space that was safe of the various phenomenological problems mentioned above and was predicted the supersymmetric spectrum and the Higgs masses. This analysis was the basis for the more detailed and updated one that will be discussed in the subsection 5.8.


\pagebreak

\noindent
\subsection{Further all loop results in softly broken supersymmetric gauge theories}
Title: Further all loop results in softly broken supersymmetric gauge theories\\
Authors: T.\ Kobayashi, J.\ Kubo, G.\ Zoupanos\\
Journal: \href{http://ac.els-cdn.com/S0370269398003438/1-s2.0-S0370269398003438-main.pdf?_tid=751b86ba-2b32-11e3-92cb-00000aab0f6c&acdnat=1380698326_af8c9619e0e04748b0433e04bdc7a68a}{Phys.\ Letts.\ {\bf B427} (1998) 291-299}\\

\noindent
Comment ({\sl Myriam Mondrag\'on, George Zoupanos}\,)\\
In this paper substantial progress has been achieved concerning the
soft supersymmetry breaking sector of $N = 1$ supersymmetric gauge theories inluding the finite ones. In
particular, the RGI sume rule discussed in subsection 5.5 up to two-loops was extended to all orders in
perturbation theory. More specifically, recalling and extending our comments on 5.5 we observe that
a RGI sum rule for the soft scalar masses exists in lower orders: it results from the independent analysis
of the SSB sector of a $N = 1$ supersymmetric GYU; in one-loop for the non-finite case \cite{Kawamura:1997cw}
and in two-loops for the finite case (subsection 5.5).
The sum rule appears to have significant phenomenological consequences and
in particular manages to overcome the unpleasant predictions of the previously
known ``universal'' finiteness condition for the soft scalar masses.

The general feeling was that hardly one could find RGI relations in the SSB sector of $N = 1$ supersymmetric theories includind the finite ones beyond
the two-loop order. However despite the negative expectations a very interesting progress has been achieved concerning the renormalization properties of
the SSB parameters. The developement was based on the powerful supergraph method for studying supersymmetric theories which has been applied to the softly broken ones by using the ``spurion'' external space-time independent superfields. According to this method a softly broken supersymmetric gauge theory is considered as a supersymmetric one in which the various parameters such as couplings and masses have been promoted to external
superfields that acquire ``vacuum expectation values''. Then based on this method certain relations among the soft term renormalization and that of an unbroken supersymmetric theory were derived. In particular the $\beta$-functions of the parameters of the softly broken theory are expressed in terms of partial differential operators involving the dimensionless
parameters of the unbroken theory. A crucial aspect  in the whole strategy for solving the set of coupled differential equations so as to be able to express all parameters in a RGI way, was to transform the partial differential
operators involved to total derivative operators. It is definitely possible to do this on the RGI surface defined by the solution of the reduction equations. 
 Using  the above tools, in the present work we proved that the sum rule for the soft scalar massses is RGI to all-orders for both the general as well as for the finite case. Finally, the exact $\beta$-function for
the soft scalar masses in the Novikov-Shifman-Vainstein-Zakharov (NSVZ) scheme for the softly broken supersymmetric QCD was obtained for the first time.
 The above method and results are of significant importance in the application of the reduction method in the MSSM and lead to important results and significant predictions, which will be discussed later in subsection 5.10.



\pagebreak


\subsection{Finite $SU(N)^k$ unification}
Title: Finite $SU(N)^k$ unification\\
Authors: E.\ Ma, M.\ Mondragon, G.\ Zoupanos\\
Journal: \href{http://iopscience.iop.org/1126-6708/2004/12/026/}{Journ.\ of High Energy Physics 0412 (2004) 026}\\

\noindent
Comment ({\sl Myriam Mondrag\'on, George Zoupanos}\,)\\
 This is a very interesting investigation since it provides the first example of a Finite Unified Theory based on gauge groups which are not simple.  The best model, which is  based on the gauge group $SU(3)^3$, is a very attractive gauge theory since being the maximal subgroup of $E_6$ it has been discussed in several investigations of GUTs, especially  in the $N = 1$ supersymmetric ones based on exceptional groups.  Moreover, it is a natural GUT obtained from the $N = 1$, 10-dimensional $E_8$ gauge group of the heterotic string theory \cite{Ibanez:2012zz,Irges:2011de} and, surpisingly, is the theory obtained in realistic four-dimensional models in which the extra dimensions are non-commutative (fuzzy) manifolds \cite{Chatzistavrakidis:2010xi}.

In the present paper we examined the possibility of constructing realistic Finite Unified Theories based on product gauge groups. In particular, we considered $N = 1$ supersymmetric theories, with gauge
groups of the type $SU(N)^1 \times SU(N)^2 \times ... \times SU(N)^k$, with $n_f$ copies (number of families) of the supersymmetric multiplets $(N,\bar{N},...,1)+(1,N,\bar{N},...,1)++...+ (\bar{N},1,1,...,N)$.
The  first and very interesting result is that a simple examination of the one-loop $\beta$-function coefficient in the renormalization group equation of each $SU(N)$ leads to the result that finiteness at one-loop requires the existence of three families of quarks and leptons for any $N$ and $k$, which also implies  that if one fixes the number of families at three the theory is automatically finite. 
Then, from phenomenological considerations an $SU(3)^3$ model is singled out. In turn an all-loop and a two-loop finite model based on this gauge group were examined and  the predictions concerning the third generation quark masses, the Higgs masses, and the supersymmetric spectrum were found. Although at the time this work was done the prediction of the top quark mass was in agreement with the corresponding experimental measurements, the latest experimental results \cite{Lancaster:2011wr} are challenging this prediction.
The same holds now for the prediction of the Higgs mass, which was found to be $\sim  130 - 132$ GeV.
There exist however ways to overcome these problems. For instance, so far in the analysis the masses of the new particles of all families appearing in the model were taken to be at the $M_{GUT}$ scale. Taking into account new thresholds for these exotic particles below $M_{GUT}$ one can hope to find a phenomenologically viable parameter space.
The details of the predictions of the $SU(3)^3$ are currently under a careful re-analysis in view of the new value of the top-quark mass, the measured Higgs mass the possible new thresholds for the exotic particles, as well as different intermediate gauge symmetry breakings.


\pagebreak


\noindent
\subsection{Confronting finite unified theories with low energy \\ phenomenology}
Title: Confronting finite unified theories with low energy  
phenomenology\\
Authors: S.\ Heinemeyer, M.\ Mondragon, G.\ Zoupanos\\
Journal: \href{http://iopscience.iop.org/1126-6708/2008/07/135}{Journ.\ of High Energy Physics 0807 (2008) 135-164}\\



Comment ({\sl Sven Heinemeyer})\\ 
After many years of theoretical preparation, finite unified theories were
ready to be confronted with phenomenology and experimental
results: the present paper is devoted to this aim. 
From the classification of theories with vanishing one-loop gauge
$\beta$ function, one can see that there exist only two candidate
possibilities to construct $SU(5)$ GUTs with three generations. These
possibilities require that the theory should contain as matter fields
the chiral supermultiplets {\bf 5, 5, 10, \boldmath{$\bar{5}$}, 24} with the
multiplicities (6, 9, 4, 1, 0) and (4, 7, 3, 0, 1), respectively. 
Only the second one contains a 24-plet which can be used to provide the
spontaneous symmetry breaking of $SU(5)$ down to 
$SU(3) \times SU(2) \times U(1)$.
The particle content of the models under consideration consists of the
following supermultiplets: three ({\bf \boldmath{$\bar{5}$} + 10}), 
needed for each of the three generations of quarks and leptons, four 
({\bf \boldmath{$\bar{5}$} + 5})  
and one {\bf \boldmath{$24$}} considered as Higgs supermultiplets. When the gauge group of
the finite GUT is broken the theory is no longer finite, and one then
assumes that one is left with the MSSM.\\
Two versions of the model were possible originally, labeled {\bf A} and 
{\bf B}. The main difference between model {\bf A} and model {\bf B} is
that two pairs of Higgs quintets and anti-quintets couple to the {\bf 24}
in {\bf B}, so that it is not necessary to mix them with $H_4$ and
$\bar{H}_4$ in order to achieve the triplet-doublet splitting after the
symmetry breaking of $SU(5)$.\\
Confronting those two models with the quark mass predictions for $m_t$
showed that only model {\bf B} can accomodate a top quark mass of about
$173 \gev$, while model {\bf A} predicted consistently $\mt \sim 183 \gev$. 
Investigating the two signs of the $\mu$ parameter revealed that only 
$\mu < 0$ predicts a bottom quark mass value in the correct range,
whereas the positive sign of $\mu$ results in $\mb$ values more than 
$1 \gev$ too high. In this way the $SU(5)$ model {\bf FUTB} was singled
out as the only phenomenological viable option.
Confronting the model predictions with the measured value of 
$\br(b \to s \ga)$ and the (then valid) upper limit on 
$\br(B_s \to \mu^+\mu^-)$ further restricted the allowed parameter
space.\\
The ``surviving'' parameter space was then used to predict the Higgs and
the SUSY spectrum to be expected in the LHC searches. The light MSSM
Higgs boson mass was predicted in a very narrow range of 
\begin{align*}
\Mh^{\rm predicted} = 121 \ldots 126 \gev~,
\end{align*}
to which a $\pm 3 \gev$ theory uncertainty has to be added.
The mass scale of the heavy Higgs bosons was predicted to be between 
$\sim 500 \gev$ and the multi-10-TeV range. The lightest observable SUSY
particle, either the light scalar tau or the second lightest neutralino,
was predicted in the range between $500 \gev$ and $\sim 4000 \gev$,
where the lighter regions was prefered by the prediction of cold dark
matter. Finally, the colored particles were predicted in the range
between $\sim 2 \tev$ and $\sim 15 \tev$, where only the lighter part of
the spectrum would allow a discovery at the LHC. 
These predictions now eagerly awaited the start of the LHC and the
experimental data on Higgs and SUSY searches.

\pagebreak


\noindent
\subsection{Finite theories after the discovery of a Higgs-like boson at the LHC}
Title: Finite theories after the discovery of a Higgs-like 
boson at the LHC\\
Authors: S.\ Heinemeyer, M.\ Mondragon, G.\ Zoupanos\\
Journal: \href{http://ac.els-cdn.com/S0370269312012956/1-s2.0-S0370269312012956-main.pdf?_tid=6978c40c-2b33-11e3-ada4-00000aacb35f&acdnat=1380698736_105ef564e2a9cbdc9dd3b8776207309a}{Phys.\ Letts.\ {\bf B718} (2013) 1430-1435}\\


Comment ({\sl Sven Heinemeyer})\\
Before the start-up of the LHC the idea of finite unified theories,
using the $SU(5)$ gauge group, resulted in only one viable
model (s.\ subsect. 5.8). Investigating the model properties yielded a clear prediction
for the Higgs and the SUSY spectrum. The light MSSM
Higgs boson mass was predicted in a very narrow range of 
\begin{align*}
  \Mh^{\rm predicted} = 121 \ldots 126 \gev~,{\hbox{\hskip 2cm} (*)}
\end{align*}
to which a $\pm 3 \gev$ theory uncertainty has to be added.
The mass scale of the heavy Higgs bosons was predicted to be between 
$\sim 500 \gev$ and the multi-10-TeV range. The lightes observable SUSY
particle, either the light scalar tau or the second lightest neutralino,
was predicted in the range between $500 \gev$ and $\sim 4000 \gev$,
where the lighter regions was prefered by the prediction of cold dark
matter. Finally, the colored particles were predicted in the range
between $\sim 2 \tev$ and $\sim 15 \tev$, where only the lighter part of
the spectrum would allow a discovery at the LHC. 
These predictions now eagerly awaited the start of the LHC and the
experimental data on Higgs and SUSY searches.\\
The spectacular discovery of a Higgs-like particle 
with a mass around $\MH \simeq 126 \gev$, which has been announced
by ATLAS \cite{Aad:2012tfa+ATLAS:2013mma} and CMS \cite{Chatrchyan:2012ufa+CMS:yva}, marks a
milestone of an effort that has been ongoing for almost half a century
and opens up a new era of particle physics.  
Both ATLAS and CMS reported a clear excess in the two photon channel, as
well as in the $ZZ^{(*)}$ channel. The discovery is further 
corroborated, though not with high significance, by the
$WW^{(*)}$ channel and by the final Tevatron results~\cite{TevHiggsfinal}.
The combined sensitivity
in each of the LHC experiments reaches more than $5\,\si$. 
Remarkably, the measured value agrees quite precisely with the value
predicted by the SU(5) finite unified theory as given in eq.\ $(*)$.
Consequently, as a crucial new ingredient one has to take into
account the recent discovery of a Higgs boson with a mass measurement of
\begin{align*}
M_h \sim 126.0 \pm 1 \pm 2 \gev~ ,
\label{eq:Mh125}
\end{align*}
where $\pm 1$ comes from the experimental error and $\pm 2$
corresponds to the theoretical error, and see how this affects the
SUSY spectrum.  Constraining the allowed values of the Higgs mass this
way puts a limit on the allowed values of the other mass parameters of
the model.  
Furthermore, no direct observation of SUSY particles has been
detected, and the lower limits on the SUSY spectrum have to be taken
into account in a realistic evaluation of the model predictions.\\
Without any $\Mh$ restrictions the
coloured SUSY particles have masses above $\sim 1.8 \tev$ in agreement
with the non-observation of those particles at the LHC. Including
the Higgs mass constraints in general favors the lower part of the
SUSY particle mass spectra, but also cuts away the very low
values. Neglecting the theory uncertainties of $\Mh$ 
permits SUSY masses which would
remain unobservable at the LHC, the ILC or CLIC.  On the other hand,
large parts of the allowed spectrum of the lighter scalar tau or the
lighter neutralinos might be accessible at CLIC with $\sqrt{s} = 3 \tev$. 
Including the theory uncertainties, even higher masses are
permitted, further weakening the discovery potential of the LHC and
future $e^+e^-$ colliders.


\pagebreak


\subsection{Reduction of Couplings in the MSSM}
Title: Reduction of Couplings in the MSSM\\ 
Authors: M.\ Mondragon, N.D.\ Tracas, G.\ Zoupanos\\ 
Journal: \href{http://ac.els-cdn.com/S0370269313009465/1-s2.0-S0370269313009465-main.pdf?_tid=963d0842-81d0-11e3-a517-00000aab0f27&acdnat=1390222042_631acd47f633ae29afd9dcd06d365009}{Phys.\ Letts.\ {\bf B728} (2014) 51-57}\\ 

\noindent
Comment ({\sl Myriam Mondrag\'on, George Zoupanos}\,)\\
This paper is of particular importance in the examination of realistic models in which the {\it reduction of couplings} can be achieved.  It is of equal theoretical importance as the paper discussed in  subsection 3.1, but more successful so far in the comparison with the known experimental facts. Moreover, contrary to the case in Finite Unified Theories, it realises the old dream of Zimmermann with asymptotic freedom at work in the reduction of the relevant couplings, as a fundamental requirement according to the original theorem.

More specifically the most important observation in this paper is that there exist RGI relations among the top, bottom Yukawa and the gauge colour couplings in the minimal supersymmetric SM, i.e. in the MSSM. This result was found by solving the reduction equations and using the power series ansatz for the solutions.  The reduced system comprises the top and bottom Yukawa couplings reduced in terms of the strong coupling, whereas the tau Yukawa coupling is left as a free parameter.  It was found that it is possible to have solutions for certain values of the tau Yukawa coupling and negative values of the $\mu$ parameter, which are consistent with the experimental results for the top and bottom quark masses simultaneously at the level of one sigma. 
Therefore the reduction of these couplings is a fact in the MSSM. Then, based on this observation and using the tools described in the subsection 5.6 it was possible to make further predictions. Assuming the existence of a RGI relation among the trilinear couplings in the superpotential and the SSB sector of the theory,  it was possible to obtain predictions for the Higgs masses and the supersymmetric spectrum. 
  It was found that the lightest Higgs mass is in the range
123.7 - 126.3 GeV, in striking agreement with the measurements at LHC \cite{Aad:2012tfa+ATLAS:2013mma,Chatrchyan:2012ufa+CMS:yva}. The rest of the spectrum  was found to be generally very heavy. Specifically, it was found that the masses of the heavier Higgses have values above the TeV scale. In addition the supersymmetric spectrum starts with a neutralino as LSP at $\sim 500$ GeV, which allows for a  comfortable agreement with the LHC bounds due to the non-observation of coloured supersymmetric particles \cite{Pravalorio:susy2012,Chatrchyan:2012vp,Campagnari:susy2012}.
 The plan is to extend the present analysis by examining the restrictions that will be imposed in the spectrum by the B-physics as well as the CDM constraints, given that the LSP in this model is in principle a candidate for CDM.

\pagebreak


\subsection{ Conclusions to Section 5}
{\sl Sven Heinemeyer, Myriam Mondrag\'on and George Zoupanos}\\
 A number of proposals and ideas have matured with time and have survived after careful theoretical
 studies and confrontation with experimental data. These include part of the original GUTs ideas, mainly
 the unification of gauge couplings and, separately, the unification of the Yukawa couplings, a version
 of fixed point behaviour of couplings, and certainly the necessity of SUSY as a way to take care of 
 the technical part of the hierarchy problem. On the other hand, a very serious theoretical problem,
 namely the presence of divergencies in Quantum Field Theories (QFT), although challenged by the founders
 of QFT \cite{Dirac:book,Dyson:1952tj,Weinberg:2009ca}, was mostly forgotten in the course of developments
 of the field partly due to the spectacular successes of renormalizable field theories, in particular of the
 SM. However, fundamental developments in theoretical particle physics are based in reconsiderations of the
 problem of divergencies and serious attempts to solve it. These include the motivation and construction 
 of string and non-commutative theories, as well as $N=4$ supersymmetric field theories
 \cite{Mandelstam:1982cb,Brink:1982wv}, $N=8$ supergravity
 \cite{Bern:2009kd,Kallosh:2009jb,Bern:2007hh,Bern:2006kd,Green:2006yu} and the AdS/CFT correspondence
 \cite{Maldacena:1997re}.  It is a thoroughly fascinating fact that many interesting ideas that have
 survived various theoretical and phenomenological tests, as well as the solution to the UV divergencies
 problem, find a common ground in the framework of $N=1$ Finite Unified Theories, which we have described
 in the previous sections. From the theoretical side they solve the problem of UV divergencies in a minimal
 way. On the phenomenological side, since they are based on the principle of reduction of couplings 
 (expressed via RGI relations among couplings and masses), they provide strict selection rules in choosing
 realistic models which lead to testable predictions.\\ 
 Currently we are still examining the predictions of the best so far $SU(5)$ Finite Unified Theory,
 including the restrictions of third generation quark masses and $B$-physics observables.  The model
 is consistent with all the phenomenological constraints. Compared to our previous analysis
 (see subsect.\ 5.8) the new bound on $\br (B_s \to \mu^+ \mu^-)$ prefers a heavier (Higgs)
 spectrum and thus in general allows only a very heavy SUSY spectrum.  The Higgs mass constraint, on
 the other hand, taking into account the improved $\Mh$ prediction for heavy scalar tops, favours the 
 lower part of this spectrum, with SUSY masses ranging from $\sim 600\gev$ up to the multi-TeV level,
 where the lower part of the spectrum could be accessible at the ILC or CLIC. Taking into account the
 improved theory uncertainty evaluation some part of the electroweak spectrum should be accessible at
 future $e^+e^-$ colliders. The coloured spectrum, on the other hand, could easily escape the LHC
 searches; also at the HL-LHC non-negligible parts of the spectrum remain beyond the discovery reach.\\
 The celebrated success of predicting the top-quark mass (see subsects. 5.1, 5.2, 5.3 and 
 \cite{Mondragon:1993tw,Kubo:1994xa,Kubo:1995zg}) has
 been extended to the predictions of the Higgs masses and the supersymmetric spectrum
 of the MSSM \cite{Heinemeyer:2007tz,Heinemeyer:2010xt}. Clear predictions for the discovery
 reach at current and future $pp$ colliders as well as for future $e^+e^-$ colliders
 result in somewhat more optimistic expectations compared to older analyses.\\


\pagebreak


\section{Discussion and Conclusions}
In the above sections we presented the historical development of two notions:
{\sl reduction of couplings} and {\sl finiteness} within $N=1$ supersymmetric
gauge theories and then how they have been applied to the standard model (SM)
and extensions of it with the aim of forcasting or describing the experimental
findings with as few parameters as possible. We selected those original papers
in which the relevant results had been obtained. These papers should speak
for themselves but by providing individual comments for them
and by putting them in the appropriate context by introductory
remarks at the beginning of the sections we tried to
make the papers and the whole endavour easier accessible
also to a reader who is not an expert in the field.\\
After having provided the machinery for reducing couplings in section
2 a first attempt to use it in particle physics has been presented
in section 3, devoted to the SM. Its final outcome in the version with
three families says that a top mass larger than roughly
111 GeV would not allow to realize asymptotic freedom of couplings
in this theory. It also shows that the results are very sensitive
to the details of the model. Already admitting a fourth generation would
change drastically the predictions. Another warning feature
came about when demanding cancellation of quadratical divergencies:
it was not very well compatible with the bound obtained for the top
mass.\\
An obvious candidate for guaranteeing absence of quadratical divergencies
related to physical parameters is supersymmetry; a way of avoiding too many new parameters is
provided by requiring finiteness. The basis for this is being given
in section 4, together with the proof that reduction is a renormalization
scheme independent concept.\\
The sequence of papers in section 5 then shows how one can
reconcile supersymmetric models with phenomenology. The first
interesting hint that this could be the right track came 
in the paper of subsection 5.1 (1992) with the prediction of 178.8 GeV
for the top mass in two finite supersymmetric $SU(5)$ models. 
At that time this has been considered as a pretty large value.\\
Encouraged by the discovery of the top around this mass value
a more systematic search has been initiated via grand unified 
supersymmetric models, unification of Yukawa couplings followed
by a careful study of supersymmetry breaking through soft
mass terms. As early as 2008 this analysis culminated eventually in
the prediction of a Higgs mass value in the interval between
$121 ... 126$ GeV (see subsect.\ 5.8). Once a Higgs-like
particle had been found experimentally its mass value
could be used for restricting further the supersymmetric
spectrum. Eventually it was possible to reproduce the experimental
value of this Higgs-like boson and to identify the lightest Higgs 
of the MSSM as {\sl the} Higgs of the standard model by partial reduction
(see subsect.\ 5.10).\\
Obviously this nice result prompts further questions. How can this
model and its renormalization group relations be linked to the
finite models which were so successful in pointing to the right
value for the top mass? Is there the respective gauge group singled out by
some specific, characterizing property? And, on top of this: Do not
all these considerations point to supersymmetry as the relevant
underlying symmetry?\\
These questions also imply that the search on the structure of matter
goes on.\\

\pagebreak

Acknowledgements\\
The present historical overview is based on the endavour of quite
a few physicists. Most of them contribute to the overview as authors. These
authors gratefully acknowledge the most valuable collaboration in the past with
Tatsuo Kobayashi, Ernest Ma and Nick Tracas.\\
In the actual course of preparing the present work the authors enjoyed,
again, helpful discussions with -- and concrete assistance by many colleagues.
In particular we are indebted to W.\ Hollik, C.\ Counnas, D.\ L\"ust,
C.\ M\~unoz, G.\ Ross, R.\ Stora and E.\ Seiler.

\noindent
The work of M.M.\ was supported by mexican grants PAPIIT IN113712 and
Conacyt 132059.\\
The work of G.Z.\ was supported by the Research Funding Program ARISTEIA,
Higher Order Calculations and Tools for High Energy Colliders, HOCTools
(cofinanced by the European Union (European Social Fund ESF) and Greek
national funds through the Operational Program Education and Lifelong
Learning of the National Strategic Reference Framework (NSRF)). G.Z.\
acknowledges also support from the European Union's ITN programme HIGGSTOOLS.\\
Accordingly G.Z.\ was affiliated to:\\
-- Max-Planck-Institut f\"ur Physik (Werner-Heisenberg-Institut)  M\"unchen, Germany\\
-- Arnold-Sommerfeld-Center f\"ur Theoretische Physik, Department f\"ur Physik,
Ludwig-Maximilians-Universit\"at, M\"unchen, Germany\\
-- Institut f\"ur Theoretische Physik, Universit\"at Heidelberg, Heidelberg, Germany\\
He thanks these Institutes for the warm hospitality extended to him.\\







\pagebreak

\providecommand{\href}[2]{#2}\begingroup\raggedright\endgroup


\begin{thebibliography}{10}

\bibitem{Hollikyellowrep}
      W.~Hollik, ``Quantum field theory and the Standard Model'',
     {\em Yellow Report CERN-2010-002, 1-44} arXiv:1012.3883[hep-ph]

\bibitem{Beringer:1900zz}
  J.~Beringer {\it et al.}  [Particle Data Group Collaboration],
  Phys.\ Rev.\ D {\bf 86} (2012) 010001.

\bibitem{Oehme_tpsuppl}
R.~Oehme, ``Reduction and reparametrization of quantum field theories''
{\em Progr.\ of Theoret.\ Phys.\ Suppl.\ No 86 (1986) 215-237}\\

\bibitem{Sibold_apprev}
K.~Sibold, ``Reduction of couplings''
{\em Acta Physica Polonica B19 (1988) 295-306}\\

\bibitem{Kraus_abBRS}
E.~Kraus, ``How abelian BRS-symmetry emerges in non-invariant surroundings''
{\em Nucl.\ Phys.\ B 349 (1991) 563-580}

\bibitem{Kraus_nonabBRS}
E.~Kraus, ``A new characerization of BRS-invariant theories with a simple
non-abelian gauge group''
{\em Nucl.\ Phys.\ B 354 (1991) 218-244}

\bibitem{Kraus_beta}
E.~Kraus, ``The $\beta$-functions of a general non-symmetric model involving
vector fields''
{\em Nucl.\ Phys.\ B 354 (1991) 245-288}

\bibitem{Piguet_alushta}
O.~Piguet, ``Supersymmetry, ultraviolet finiteness and grand unification'';
Talk given at Conference: C96-05-13;
e-Print: hep-th/9606045

\bibitem{Kraus_susy_anom1}
E.~Kraus, ``An anomalous breaking of supersymmetry in supersymmetric
gauge theories with local coupling''
{\em Nucl.\ Phys.\ B 620 (2002) 55-83}


\bibitem{Kraus_susy_anom2}
E.~Kraus, ``Calculating the anomalous supersymmetry breaking
in Super-Yang-Mills with local coupling''
{\em Phys.\ Rev.\ D65 (2002) 105003}

\bibitem{KrausRuppSibold}
E.~Kraus, C.~Rupp, K.~Sibold ``Supersymmetric Yang-Mills theories
with local coupling: The supersymmetric gauge''
{\em Nucl.\ Phys.\ B 661 (2003) 83-98}

\bibitem{Tseytlin}
A.A.~Teytlin, ``On partition function and Weyl anomaly of conformal higher
spin fields''
arXiv:1309.0785v4 [hep-th] 28 Oct 2013\\

\bibitem{Lancaster:2011wr}
 Tevatron Electroweak Working Group, CDF and D0 Collaborations,
 \newblock (2011), 1107.5255.
 \bibitem{Aad:2012tfa+ATLAS:2013mma}
 ATLAS Collaboration, G. Aad et~al.,
 \newblock Phys.Lett. B716 (2012) 1, 1207.7214;
 %
 ATLAS Collaboration, Reports ATLAS-CONF-2013-014,
 ATLAS-COM-CONF-2013-025
 \newblock (2013).
 
 \bibitem{Chatrchyan:2012ufa+CMS:yva}
 CMS Collaboration, S. Chatrchyan et~al.,
 \newblock Phys.Lett. B716 (2012) 30, 1207.7235;
 %
 CMS Collaboration, Report CMS-PAS-HIG-13-005
 \newblock (2013).
 
 \bibitem{Mondragon:2013aea}
 M. Mondragon, N. Tracas and G. Zoupanos,
 \newblock accepted in Phys. Lett. B  (2013), 1309.0996.
 
\bibitem{Kachru:1998ys}
  S. Kachru and E. Silverstein,
 \newblock Phys. Rev. Lett. 80 (1998) 4855, hep-th/9802183.
 
 \bibitem{Chatzistavrakidis:2010xi}
 A. Chatzistavrakidis, H. Steinacker and G. Zoupanos,
 \newblock JHEP 1005 (2010) 100, 1002.2606.
 
 \bibitem{Heinemeyer:2011aa}
 S. Heinemeyer, O. Stal and G. Weiglein,
 \newblock Phys.Lett. B710 (2012) 201, 1112.3026.
 
 \bibitem{Kubo:1995cg}
 J. Kubo et~al.,
 \newblock Nucl. Phys. B479 (1996) 25, hep-ph/9512435.
 
 \bibitem{Kubo:1994xa}
 J. Kubo et~al.,
 \newblock Phys. Lett. B342 (1995) 155, hep-th/9409003.



\bibitem{ATLASdiscovery} G.~Aad {\it et al.}  [ATLAS Collaboration],
Phys.\ Lett.\ B {\bf 716} (2012) 1
[arXiv:1207.7214 [hep-ex]].

\bibitem{CMSdiscovery} S.~Chatrchyan {\it et al.}  [CMS Collaboration],
Phys.\ Lett.\ B {\bf 716} (2012) 30
[arXiv:1207.7235 [hep-ex]].

\bibitem{TevHiggsfinal} CDF Collaboration, D\O Collaboration,
[arXiv:1207.0449 [hep-ex]].



\bibitem{Kubo:1995zg}
 J. Kubo et~al.,
 \newblock Nucl. Phys. B469 (1996) 3, hep-ph/9512258.
 
 \bibitem{Kawamura:1997cw}
 Y. Kawamura, T. Kobayashi and J. Kubo,
 \newblock Phys. Lett. B405 (1997) 64, hep-ph/9703320.
 
 \bibitem{Ibanez:2012zz}
 L.E. Ibanez and A.M. Uranga,
 \newblock {String theory and particle physics: An introduction to string
   phenomenology} (Cambridge University Press, 2012).
 
 \bibitem{Irges:2011de}
 N. Irges and G. Zoupanos,
 \newblock Phys.Lett. B698 (2011) 146, 1102.2220.
 
 \bibitem{Pravalorio:susy2012}
 ATLAS Collaboration, P. Pravalorio,
 \newblock Talk at SUSY2012  (2012).
 
 \bibitem{Chatrchyan:2012vp}
 CMS Collaboration, S. Chatrchyan et~al.,
 \newblock Phys.Lett. B713 (2012) 68, 1202.4083.
 
 \bibitem{Campagnari:susy2012}
 CMS Collaboration, C. Campagnari,
 \newblock Talk at SUSY2012  (2012).
 
\bibitem{Dirac:book}
  P. Dirac,
  \newblock {Lectures On Quantum Field Theory} .
	
 \bibitem{Dyson:1952tj}
   F.J. Dyson,
 \newblock Phys. Rev. 85 (1952) 631.
 
 \bibitem{Weinberg:2009ca}
   S. Weinberg,
   \newblock (2009), 0903.0568,
    \newblock and references therein.
 
  \bibitem{Mandelstam:1982cb}
   S. Mandelstam,
  \newblock Nucl. Phys. B213 (1983) 149.
 



\bibitem{Brink:1982wv}
    L. Brink, O. Lindgren and B.E.W. Nilsson,
  \newblock Phys. Lett. B123 (1983) 323.
 
\bibitem{Bern:2009kd}
    Z. Bern et~al.,
  \newblock Phys. Rev. Lett. 103 (2009) 081301, 0905.2326.
  
\bibitem{Kallosh:2009jb}
    R. Kallosh,
  \newblock JHEP 09 (2009) 116, 0906.3495.
 
\bibitem{Bern:2007hh}
    Z. Bern et~al.,
   \newblock Phys. Rev. Lett. 98 (2007) 161303, hep-th/0702112.
 
 \bibitem{Bern:2006kd}
     Z. Bern, L.J. Dixon and R. Roiban,
   \newblock Phys. Lett. B644 (2007) 265, hep-th/0611086.
 
 \bibitem{Green:2006yu}
      M.B. Green, J.G. Russo and P. Vanhove,
 \newblock Phys. Rev. Lett. 98 (2007) 131602, hep-th/0611273.
 
 \bibitem{Maldacena:1997re}
      J.M. Maldacena,
  \newblock Adv. Theor. Math. Phys. 2 (1998) 231, hep-th/9711200.
 
 \bibitem{Heinemeyer:2007tz}
      S. Heinemeyer, M. Mondragon and G. Zoupanos,
 \newblock JHEP 07 (2008) 135, 0712.3630.
 
 \bibitem{Kapetanakis:1992vx}
     D. Kapetanakis, M. Mondragon and G. Zoupanos,
    \newblock Z. Phys. C60 (1993) 181, hep-ph/9210218.
  
  \bibitem{Mondragon:1993tw}
      M. Mondragon and G. Zoupanos,
     \newblock Nucl. Phys. Proc. Suppl. 37C (1995) 98.
 
  \bibitem{Kubo:1994bj}
      J. Kubo, M. Mondragon and G. Zoupanos,
    \newblock Nucl. Phys. B424 (1994) 291.
 
 \bibitem{Kubo:1996js}
         J. Kubo, M. Mondragon and G. Zoupanos,
	 \newblock Phys. Lett. B389 (1996) 523, hep-ph/9609218.

 \bibitem{Heinemeyer:2010xt}
        S. Heinemeyer, M. Mondragon and G. Zoupanos,
      \newblock SIGMA 6 (2010) 049, 1001.0428.




\end{thebibliography}
\end{document}